\documentclass{elsarticle}

\usepackage[utf8]{inputenc}
\usepackage{amsmath}
\usepackage{amsfonts}
\usepackage{amssymb}
\usepackage{color}
\usepackage{graphicx}
\usepackage{siunitx}
\usepackage[capitalize]{cleveref} 
\usepackage[version=4]{mhchem}
\usepackage{float}
\usepackage{subfig}
\usepackage{mathrsfs}
\usepackage{comment}
\usepackage{upgreek}
\usepackage{tikz}
\usepackage{gensymb}
\usetikzlibrary{arrows,backgrounds,patterns,shapes.geometric,calc}
\usetikzlibrary{positioning}

\graphicspath{{./figures/}} 

\biboptions{sort&compress}

\def\doubleunderline#1{\underline{\underline{#1}}}
\def\tensor#1{\doubleunderline{\mathbf{#1}}}
\def\greektensor#1{\doubleunderline{\boldsymbol{#1}}}
\def\ii{{\mathrm i}}

\begin{document}
\begin{frontmatter}

\title{On the scattering of
entropy waves at sudden area expansions}

\author[inst1]{Juan Guzm\'an-I\~nigo} 
\author[inst2]{Dong Yang}
\author[inst1]{Renaud Gaudron}
\author[inst1]{Aimee S. Morgans} 

\address[inst1]{Department of Mechanical Engineering, Imperial College London, London, UK}
\address[inst2]{Department of Mechanics and Aerospace Engineering, Southern University of Science and Technology, Shenzhen, China}

\begin{abstract}
In this work, we investigate both numerically and theoretically
the sound generated by entropy waves passing through 
sudden area expansions. This is a canonical configuration 
representing internal flows with flow separation 
and stagnation pressure losses. The numerical approach is based on a triple 
decomposition of the flow variables into 
a steady mean, a small-amplitude 
coherent part, and a stochastic turbulent part. The coherent 
part contains acoustic, vortical, and entropy  waves. The mean
flow is obtained as the solution of the 
Reynolds-Averaged Navier-Stokes (RANS) equations. The 
equations governing the coherent perturbations are linearised 
and solved in the frequency domain. To account for the effect
of turbulence on the coherent perturbations, a frozen eddy viscosity 
model is employed. When entropy fluctuations pass through the area 
expansion, the generated entropy noise behaves 
as a low-pass filter. The numerical predictions of the noise at low 
frequencies are compared to the predictions of
compact, quasi-one-dimensional, and isentropic theory and large
discrepancies are observed. An alternative model 
for the generated entropy noise tailored for area
 expansions is then proposed. Such model 
is based on the conservation of mass, momentum, and 
energy written in integral form. The model assumes zero frequency and
the one-dimensionality of the flow variables far upstream and downstream 
of the expansion. The predictions of this model agree well  
with the numerical simulations across a range of finite 
subsonic Mach numbers including low, intermediate, and high Mach numbers.
The contributions of this work are both numerical and theoretical.
Numerically, a triple decomposition 
adapted to high-Mach-number, compressible flows is introduced 
for the first time in the context of acoustic simulations. From 
a theoretical point of view, 
the model proposed here represents the first quasi-steady model that 
captures correctly the low-frequency entropy noise generated at 
sudden area expansions, including at high subsonic Mach numbers.
\end{abstract}

\end{frontmatter}

\section{Introduction}\label{sec:introduction}


Small-amplitude perturbations superimposed on
a single-component, compressible flow can be classified 
into three types of disturbances~\cite{kovasznay1953turbulence,chu58}:
(i) acoustic waves that are homentropic and irrotational,
(ii) entropy waves that are incompressible and irrotational, and
(iii) vortical waves that are incompressible and homentropic. 
If the flow is uniform, and viscous effects and 
heat transfer are negligible, these waves are independent
and can be considered separately. In this case, acoustic waves 
propagate in all directions in relation 
to the flow at the speed of sound, while entropy 
and vortical waves are simply swept by the 
flow silently.

Viscous terms, heat transfer, mean-flow gradients, and solid 
boundaries can act as coupling terms. 
For instance, acoustic waves can be scattered into vortical 
waves by sharp edges~\cite{crighton1985kutta} or dissipated
into heat (entropy) by viscous effects. Likewise, entropy
waves can generate sound when accelerated/decelerated
(mean-flow gradients)~\cite{marble77,bake2009entropy} or 
when interacting with solid boundaries, such as 
aerofoils~\cite{guzman2018model,guzman2019noise,guzman2020entropy}.
This is termed entropy noise and 
is particularly relevant in the context of 
confined combustion, as occurs in the combustors of 
gas turbines~\cite{morgans16}
and rockets engines~\cite{keller1985thermally}.


Unsteady combustion is, through unsteady heat release rate, 
a source of flow disturbances. In gas turbines, these 
disturbances are the origin of two distinctive sources
of noise~\cite{dowling15,ihme2017combustion,tam2019combustion}:
(i) direct and (ii) indirect sources.
Unsteady heat release rate, through volumetric 
expansion~\cite{thomas1966flame}, acts as a monopole source 
of acoustic fluctuations, the so-called direct 
combustion noise.
Unsteady combustion also generates advective disturbances  
in the form of vorticity, temperature (entropy), 
and mixture-composition fluctuations. These are silent while 
advecting in the combustor but generate noise at the exhaust
when interacting with the 
turbine blade rows~\cite{cumpsty77}. This source of noise
is termed indirect combustion noise and can be 
further decomposed as
vortical~\cite{kings2010indirect,hirschberg2021sound,hirschberg2021swirl}, 
entropy~\cite{marble77, bake2009entropy} and, 
compositional 
noise~\cite{strahle1976noise,magri2016compositional,de2021compositional}
after the name of the flow perturbation which creates them.
Entropy noise is believed to be its main 
component~\cite{morgans16}.



Entropy noise produced by the 
acceleration of entropy waves in nozzle flows
was demonstrated experimentally by the
Entropy Wave Generator (EWG) test rig at the German 
Aerospace Center (DLR)~\cite{bake2009entropy}. 
Leyko et al.~\cite{leyko2011numerical} showed that 
the experimental results for sonic throat conditions can
be reproduced using the compact theory of 
Marble and Candel~\cite{marble77}. This seminal work
proposed an analytical solution 
of the linearised, quasi-one-dimensional, and isentropic Euler equations.
The model assumes zero frequency and is valid for
configurations where the wavelengths of both the entropy 
and acoustic waves are long compared to the length 
of the nozzle (the compact assumption). 

Duran et al.~\cite{duran2013analytical} 
also compared the DLR EWG experimental results with the
compact theory~\cite{marble77}, in this case for 
subsonic conditions. They found that the model qualitatively 
predicts the shape of the experimental signals, but 
quantitative disagreement was observed. The
mismatch was attributed to the non-compactness of 
the entropy waves for this condition owing to the lower 
convective velocity. Several 
efforts have been made to extend the frequency range 
of validity of the compact theory, including analytical 
solutions for nozzles with linear velocity 
profiles~\cite{moase2007forced,giauque2012analytical} and
the use of asymptotic expansions of the governing 
equations in terms of frequency~\cite{stow02,goh11}.
Duran and Moreau~\cite{duran13magnus} proposed a solution 
based on the Magnus expansion method that is valid at any 
frequency within the range of validity of the quasi-one-dimensional
assumption. Above a low-frequency limit, the entropy 
profiles become three-dimensional. This effect must be included 
in the equations of the acoustics that can be 
treated as one-dimensional~\cite{emmanuelli2020description,huet2020entropy}. 
Further extensions of the quasi-one-dimensional theory include
non-linear effects~\citep{huet13}, compositional 
inhomogeneites~\citep{magri2016compositional,magri2017indirect}, 
mean-flow heat trasfer~\cite{yeddula2021heat}, and viscous 
effects~\cite{huet2021viscosity}. 

A limitation of the experimental campaign of the DLR EWG 
was the lack of measurements of entropy noise upstream of the 
nozzle (the so-called reflection coefficient). The study of this 
component is of great technical importance because it can modify 
the stability of the combustor~\cite{polifke2001constructive,goh13} 
leading to large-amplitude, self-sustained thermoacoustic 
instabilities~\cite{candel02, lieuwen03}. To address this 
limitation, a new rig was designed 
at the University of Cambridge, the Cambridge Wave
Generator (CWG)~\cite{de2017detection}, where the entropy spots 
were accelerated through an orifice plate. Rolland 
et al.~\cite{rolland2017theory} compared the experimental 
results with the predictions of quasi-one-dimensional 
and isentropic models~\cite{marble77,duran13magnus}, showing that 
the experimental reflection coefficient under subsonic conditions 
was significantly higher than predicted by such theories. The difference was
attributed to the intrinsic differences between an isentropic nozzle, 
as described by the aforementioned theories, and an orifice plate, as 
used in the experiment. The orifice comprises 
a sudden area contraction followed by an expansion. 
In the contraction portion, the flow remains essentially 
attached and isentropic. However, in the expansion 
part, the flow separates, creating a large recirculation zone which leads to 
turbulent friction and strong 
mean-entropy gradients at the downstream side. 
De Domenico et al.~\cite{de2019generalised,de2021compositional}
proposed the mean non-isentropicity produced at the area expansion
as the source of the discrepancy 
and derived an alternative compact model 
that successfully reproduces the experimental results. 
 
To obtain further physical insight into the entropy-noise generation 
mechanism in non-isentropic flows, Yang 
et al.~\cite{yang2020entropy} theoretically studied a sudden area
expansion using an acoustic analogy. They found that,
for low-Mach-number flows, the quasi-one-dimesional 
and isentropic theory of Marble and Candel~\cite{marble77} 
significantly overpredicts the actual reflection 
coefficient generated by entropy deceleration. The physical origin
of the disagreement was explained in isentropic terms: 
the existence and length of the recirculation region creates a
spatial gap between the source region and the geometrical area expansion
and this produces a different acoustic response from when 
both are aligned. When the source term and the area expansion are 
aligned, the reflection coefficient was found to tend to the 
predictions of Marble and Candel~\cite{marble77} at low 
frequencies. The authors 
concluded that the main factor for the 
inability of the Marble and Candel model to capture the entropic response 
of sudden area expansions sustaining low-Mach-number 
bias flows is in fact the spatial delay of the source term 
associated with separation, and 
not the non-isentropicity of the flow as previously suggested.

The theoretical analysis of Yang et al.~\cite{yang2020entropy} 
was, however, restricted to low-Mach-number flows. In this work, 
we investigate both theoretically and numerically 
the sound generated by entropy fluctuations decelerated at
sudden area expansions for several 
Mach numbers, including high subsonic Mach numbers. The 
theoretical model proposed is an extension of the zero-frequency model 
developed by Ronneberger~\cite{ronneberger1967,ronneberger1987}
for the acoustic scattering of area expansions. Here, 
we extend it to account for incoming entropy fluctuations. The 
approach only requires one-dimensionality of the flow variables
at the inlet and at the outlet and places 
no dimensional restrictions in the close vicinity of the
expansion itself.    

The numerical approach is based on a linearisation
of the Navier-Stokes equations 
about a turbulent mean flow. The linearised equations are then 
solved in the frequency domain to avoid spurious wave 
solutions that appear in the time domain~\cite{rao2006use}. A similar 
approach was used to successfully describe the acoustic response of 
perforations~\cite{kierkegaard2010frequency} and
sudden area expansions~\cite{kierkegaard2011simulations,gikadi2012linearized}
sustaining low-Mach-number bias flows. In these publications, two simplifications 
were considered: (i) the effect of turbulence on the acoustic field was 
neglected and (ii) the resolution of the energy equation was avoided 
by assuming that the mean flow was incompressible and the acoustics isentropic.
The latter assumption is justified owe to the low Mach number of the 
configurations of interest. The first simplification was relaxed 
by~\cite{gikadi2014impact, holmberg2015frequency} by 
decomposing the flow variables into mean, coherent 
and, turbulent parts (the so-called triple decomposition)
as proposed by Reynolds and Hussain~\cite{reynolds1972mechanics}.
Additionally, some authors have considered the energy equation for 
laminar flows, e.g. Schulze et al.~\cite{schulze2016linearized} 
to characterise perforations and
Ullrich et al.~\cite{ullrich2014acoustic}
to investigate the acoustic-entropic coupling in a 
Laval nozzle. In this work, we consider both the effect of turbulence 
and the energy equation in the governing equations. To this end, we
adapt the triple decomposition for compressible flows proposed
by Alizard et al.~\cite{alizard2015optimal} to the context of acoustic 
simulations.

The contributions of this paper are threefold. First, we present, for the first 
time in the context of acoustic simulations, a triple decomposition adapted to
high-Mach-number, compressible flows. Secondly, we adapt 
a compact theory tailored for area expansions to account 
for incoming entropy fluctuations. Thirdly, we use these
two approaches to systematically characterise the unsteady 
response of a subsonic area expansion forced by incoming 
entropy fluctuations, a configuration that represents a 
canonical example of internal flows with large-scale separation and  
stagnation pressure losses (non-isentropicity).

This article is structured as follows. First, the configuration of interest
is introduced in Section~\ref{sec:Problem formulation}. Secondly, the 
numerical and analytical approaches
are presented in Sections~\ref{sec:numerical method} 
and~\ref{sec:analytical modelling}, respectively. Results are then given 
in Sections~\ref{sec:mean flow} and~\ref{sec:entropy} for the mean flow 
and the perturbation response, respectively. Finally, conclusions are drawn
in Section~\ref{sec:conclusion}. 

\section{Problem formulation}\label{sec:Problem formulation} 
\begin{figure}
  \begin{tikzpicture}

\tikzset{
    partial ellipse/.style args={#1:#2:#3}{
        insert path={+ (#1:#3) arc (#1:#2:#3)}
    }
}

\useasboundingbox (0.0,0.0) rectangle (12.0, 3);


\draw[line width=1] (2.2, 1.0) -- (5.5, 1.0);
\draw[line width=1] (2.2, 2.0) -- (5.5, 2.0);

\draw[line width=0.5,dashed] (5.0, 1.0) -- (6.0, 1.0);
\draw[line width=0.5,dashed] (5.0, 2.0) -- (6.0, 2.0);

\draw [line width=1] (2.2, 1.5) [partial ellipse=-90:-270:0.3 and 0.5];
\draw [line width=0.5,dashed] (2.2, 1.5) [partial ellipse=-90:90:0.3 and 0.5];

\shade[top color=blue!5!white, bottom color=blue!60!white,opacity=0.3] (2.2, 1.5) [partial ellipse=-90:-270:0.3 and 0.5] -- (4.0,2.0) -- (4.0, 1.0)-- cycle;
\shade[top color=blue!5!white, bottom color=blue!60!white,opacity=0.3] (5.6, 1.5) [partial ellipse=-90:-270:0.24 and 0.5] -- (4.0,2.0) -- (4.0, 1.0)-- cycle;

\draw[line width=0.6, blue, smooth,variable=\x,samples=200, yshift=1.25cm, domain =3.7:4.2] plot ({\x},{0.1*sin(1440*(\x-0.2))});
\draw[line width=0.6, blue, ->] (3.7, 1.25) -- (3.6, 1.25);
\node at (4.5, 1.25) {{\color{blue} $\hat{p}_u^-$}};

\draw[line width=0.6, blue, smooth,variable=\x,samples=200, yshift=1.75cm, domain =4.15:4.65] plot ({\x},{0.1*sin(1440*(\x-.15))});
\draw[line width=0.6, blue, ->] (4.65, 1.75) -- (4.75, 1.75);
\node at (3.9, 1.75) {{\color{blue} $\hat{p}_u^+$}};

\draw[line width=0.6, red, smooth,variable=\x,samples=200, yshift=1.5cm, domain =3:3.5] plot ({\x},{0.1*sin(1440*(\x))});
\draw[line width=0.6, red, ->] (3.5, 1.5) -- (3.6, 1.5);
\node at (2.8, 1.5) {{\color{red} $\hat{s}^+_u$}};

\draw[line width=0.6, ->] (1.2, 1.7) -- (1.5, 1.7);
\draw[line width=0.6, ->] (1.2, 1.5) -- (1.5, 1.5);
\draw[line width=0.6, ->] (1.2, 1.3) -- (1.5, 1.3);
\node at (1.6, 2.0) {$\bar{u}_u$};

\node at (2.2, 1.5) {$A_u$};
\node at (6.0, 2.3) {$A_b$};


\draw [line width=1] (6.0, 1.5) [partial ellipse=-90:-270:0.6 and 1.2];
\draw [line width=0.5,dashed] (6.0, 1.5) [partial ellipse=-90:90:0.6 and 1.2];
\draw [line width=0.5,dashed] (6.0, 1.5) ellipse (0.3 and 0.5);

\draw[line width=1] (6.0, 0.3) -- (10.0, 0.3);
\draw[line width=1] (6.0, 2.7) -- (10.0, 2.7);
\draw[line width=1] (10.0, 1.5) ellipse (0.6 and 1.2);

\shade[top color=blue!5!white, bottom color=blue!60!white,opacity=0.3] (6.0, 1.5) [partial ellipse=-90:-270:0.6 and 1.2] --  (8, 2.7) -- (8, 0.3) -- cycle;
\shade[top color=blue!5!white, bottom color=blue!60!white,opacity=0.3] (10.0, 1.5) [partial ellipse=90:-90:0.6 and 1.2] -- (8.0, 0.3) -- (8.0, 2.7) -- cycle;

\draw[line width=0.6, blue, smooth,variable=\x,samples=200, yshift=2.0cm, domain =8.5:8] plot ({\x},{0.1*sin(1440*(\x))});
\draw[line width=0.6, blue, ->] (8.5, 2.0) -- (8.6, 2.0);
\node at (7.7, 2.0) {{\color{blue} $\hat{p}_d^+$}};

\draw[line width=0.6, blue, smooth,variable=\x,samples=200, yshift=1.0cm, domain =8.5:8] plot ({\x},{0.1*sin(1440*(\x))});
\draw[line width=0.6, blue, ->] (8.0, 1.0) -- (7.9, 1.0);
\node at (8.9, 1.0) {\color{blue} $\hat{p}_d^-$};

\draw[line width=0.6, ->] (10.8, 1.8) -- (11.0, 1.8);
\draw[line width=0.6, ->] (10.8, 1.5) -- (11.0, 1.5);
\draw[line width=0.6, ->] (10.8, 1.2) -- (11.0, 1.2);
\node at (11.1, 2.1) {$\overline{u}_d$};

\node at (10.0, 1.5) {$A_d$};

\end{tikzpicture}
  \caption{Schematic of the configuration. The subindices
$(\cdot)_u$ and $(\cdot)_d$ denote variables defined in the upstream and
downstream ducts, respectively. $A_b$ denotes the area of the backplate.}
  \label{fig:Configuration diagram}
\end{figure}
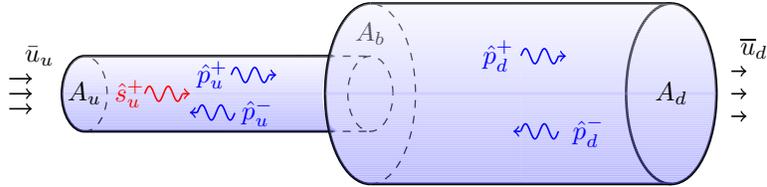
We consider a canonical area expansion consisting of two 
concentric cylinders with different cross-sectional areas as 
sketched in figure~\ref{fig:Configuration diagram}.
A subsonic, uniform flow in the streamwise direction is 
imposed on the left-hand side of the domain. The 
left-hand- and right-hand-side ducts are denoted, respectively, 
as upstream and downstream ducts and all the variables
defined in them are denoted by the subscripts $(\cdot)_u$ 
and $(\cdot)_d$, respectively.
The expansion ratio, $\eta=A_u/A_d,$
defined as the ratio of the cross-sectional areas of the 
upstream, $A_u,$ and downstream ducts, $A_d,$ 
fully determines the geometry of the problem.
The backplate, defined as the portion of the 
external wall located where the upstream 
duct expands, has an area $A_b=A_d-A_u.$ The
variables defined on the backplate are denoted by 
the subscript $(\cdot)_b.$ 

The geometry exhibits cylindrical symmetry around the central axis
and, thus, can be described using a cylindrical 
system of coordinates centred in the axis at the location of 
the expansion. The axial, radial, and azimuthal coordinates
are denoted by $x,$ $r,$ and $\theta,$ respectively, and 
its unit vectors by $\mathbf{e}_x,$ $\mathbf{e}_r,$ 
and $\mathbf{e}_{\theta},$ respectively. 
Additionally, we assume that the flow variables
are axisymmetric.

We assume that the flow at the inlet consists of a uniform component of 
velocity $\bar{\mathbf{u}}_u=\bar{u}_u \mathbf{e}_x,$ 
density $\bar{\rho}_u$, and speed of sound $\bar{c}_u$,
on which there is superimposed a small unsteady motion. Consequently, 
the flow variables can be decomposed into a steady mean, denoted by $\bar{(\cdot)}$,
and a perturbation component, denoted by $\tilde{(\cdot)}$. 
The mean flow velocity is normalised by the speed of sound 
to define the Mach number, i.e. $M=|\bar{\mathbf{u}}|/\bar{c}.$
The perturbation component is 
small and, thus, the equations governing its evolution can be linearised around the
mean flow allowing us to use the harmonic ansatz 
\begin{equation}
\tilde{(\cdot)}=\hat{\left(\cdot\right)}\exp\left(\ii\omega t\right),
\label{eq:harmonic ansatz}
\end{equation}
where $\omega$ is the angular frequency, $t$ the time, and $\ii$ the imaginary unit.
The angular frequency can be normalised by the diameter of 
the upstream duct, $D_u,$ and by either the mean velocity in the 
upstream duct, $\bar{u}_u$, to
define a Strouhal number  
\begin{equation}
St=\frac{\omega D_u}{\bar{u}_u},
\label{eq:Strouhal number}
\end{equation}
or by the speed of sound in the upstream duct, $\bar{c}_u,$
to define a Helmholtz number
 \begin{equation}
He=\frac{\omega D_u}{\bar{c}_u}.
\label{eq:Helmholtz number}
\end{equation}
Far from the area expansion, the acoustic field can be assumed to be 
composed of plane waves propagating in the same and opposite direction 
of the mean flow and denoted by $\hat{\left(\cdot\right)}^+$ 
and $\hat{\left(\cdot\right)}^-,$ respectively. The perturbation 
pressure and axial velocity can be written in terms of 
the plane waves as
$\hat{p}=\hat{p}^++\hat{p}^-$ and 
$\hat{u}=\left(\hat{p}^+-\hat{p}^-\right)/\left(\bar{\rho}\bar{c}\right),$ respectively.


The plane acoustic waves can be classified into
incoming waves, which are inputs to the system, and  outgoing 
waves, which are outputs. Here, we consider 
uniform entropy waves imposed at the inlet, $\hat{s}_u^+,$ as the input,
with the two outputs being the reflected, $\hat{p}_u^-,$ and
transmitted, $\hat{p}_d^+$, acoustic waves. The incoming entropy 
fluctuations are normalised by the specific heat at 
constant pressure $c_p,$ such that $\sigma=\hat{s}_u^+/c_p.$ The 
two outgoing acoustic waves are normalised as 
\begin{equation}
P^-_u=\hat{p}_u^-/\gamma\bar{p}_u,\quad
P^+_d=\hat{p}_d^+/\gamma\bar{p}_d,
\label{eq:coefficients}
\end{equation} 
where $\gamma$ denotes the adiabatic index. The reflection
and transmission coefficients are then defined as
$P^-_u/\sigma$ and $P^+_d/\sigma,$ respectively. 
In this work, we seek to retrieve these 
reflection and transmission coefficients
using two different techniques: 
a numerical approach, presented in Sec.~\ref{sec:numerical method}, and an 
analytical model, presented in Sec.~\ref{sec:analytical modelling}.

\section{Numerical modelling}\label{sec:numerical method}
In this section, we present the numerical approach used to 
characterise the acoustic response of the sudden area 
expansion. The derivation of the governing equations
starts from the compressible Navier-Stokes
equations. A triple decomposition
is then introduced (Sec.~\ref{sec:governing equations}) and equations 
for the mean and perturbation parts are derived in Sec.~\ref{sec:mean flow numerical}  
 and Sec.~\ref{sec:acoustic numerical}, respectively.
 
\subsection{Governing equations}\label{sec:governing equations}
The working fluid is taken to be a compressible, calorically perfect gas. 
The equations describing the flow
are the conservation of mass, momentum, and energy that read, respectively,
\begin{subequations}
\begin{align}
\frac{\partial \rho}{\partial t} + \nabla\cdot\left({\rho \mathbf{u}}\right) &= 0, \\
\frac{\partial \left( \rho \mathbf{u} \right)}{\partial t} 
+ \nabla \cdot \left( \rho \mathbf{u} \otimes \mathbf{u}\right) &= -\nabla p + \nabla \cdot\greektensor{\uptau}, \\
\frac{\partial \left( \rho e_{t}  \right) }{\partial t} 
+ \nabla \cdot \left( \rho \mathbf{u}e_t + p\mathbf{u} \right) &= \nabla\cdot\left(\mathbf{u}\cdot\greektensor{\uptau}\right) - \nabla \cdot \mathbf{q},
\label{eq:Compressible_NS_Energy}
\end{align}
\label{eq:Compressible_NS}
\end{subequations}
where $\rho$ is the density, $\mathbf{u}$ the velocity, $p$ the pressure, and $e_t$ the 
total energy. Heat radiation is neglected. The fluid is taken to 
be Newtonian and, thus, the viscous stress tensor, $\greektensor{\uptau},$ is given by
\begin{equation}
\greektensor{\uptau} = \mu \left[ \left(\nabla \mathbf{u} 
+ \nabla \mathbf{u}^{\top}\right) 
- \frac{2}{3}\left(\nabla \cdot \mathbf{u}\right)\greektensor{\updelta}\right],
\label{eq:viscous tensor}
\end{equation}
where $\mu$ is the molecular dynamic viscosity and $\greektensor{\updelta}$ is 
the Kronecker delta unit tensor. Using the 
Fourier law, the heat-conduction vector, $\mathbf{q},$ is defined  as
\begin{equation}
\mathbf{q}=-\mu\frac{c_p}{Pr}\nabla T,
\label{eq:heat transfer}
\end{equation}
where $T$ is the temperature and $Pr$ the Prandtl number.
The temperature is related to the total energy 
by the relation $e_t=c_v T + \mathbf{u}\cdot \mathbf{u}/2,$ where $c_v$ is 
the specific heat at constant volume. Additionally, the equation of state
relates temperature, density and pressure as $p = \rho R T,$ with $R$
the specific gas constant. 

As explained in Sec.~\ref{sec:Problem formulation}, the flow 
variables are decomposed into mean and perturbation parts. Following
Reynolds and Hussain~\cite{reynolds1972mechanics}, such 
decomposition can be extended to account for turbulent 
motions using the so-called 
triple decomposition. The triple decomposition of the density and 
pressure fields into mean flow, coherent perturbations,
and small-scale random motions reads
\begin{equation}
\rho = \bar{\rho} + \tilde{\rho} + \rho' \qquad \text{and} 
\qquad p = \bar{p} + \tilde{p} + p',
\label{eq:triple decomposition}
\end{equation}
with the definition of each component given by
\begin{equation}
\bar{f}(\mathbf{x}) = \lim_{L\rightarrow\infty} \frac{1}{L}\int^{L}_0 f(\mathbf{x}, t){\rm d}t,  \qquad \tilde{f}= \left\langle f \right\rangle - \bar{f}, \qquad f' = f -  \left\langle f \right\rangle.
\label{eq:triple decomposition definition}
\end{equation}
The phase-average operator, $\left\langle\cdot\right\rangle$, is defined by
\begin{equation}
 \left\langle f \right\rangle = \lim_{N\rightarrow\infty} \frac{1}{N}\sum^{N}_{n=0} f(\mathbf{x}, t+ nT_{\left\langle\right\rangle}), 
\label{eq:Phase average}
\end{equation}
with $T_{\left\langle\right\rangle}$ the period of the harmonic motion, i.e $T_{\left\langle\right\rangle}=2\pi/\omega$. Note that the coherent component contains the
acoustic motions, as well as vortical and entropic fluctuations.

When considering compressible flows, it is convenient to introduce a 
density-weighted triple decomposition as proposed by 
Alizard et al.~\cite{alizard2015optimal}. This decomposition is applied 
to the velocity, $\mathbf{u}$, total energy, $e_t$, 
and temperature, $T$, fields, such that
\begin{equation}
\mathbf{u} = \bar{\mathbf{u}}_{\rm f} + \tilde{\mathbf{u}}_{\rm f} + \mathbf{u}'_{\rm f}, 
\label{eq:Favre triple decomposition}
\end{equation}
 where the subindex $(\cdot)_{{\rm f}}$ denotes density-weighted components (also 
 called Favre-averaged components). The definitions of these terms are
\begin{equation}
\bar{f}_{\rm f} = \frac{\overline{\rho f}}{\bar{\rho}},  \qquad \tilde{f}_{\rm f}= \frac{\left\langle \rho f \right\rangle}{\left\langle \rho \right\rangle} - \bar{f}_{\rm f}, \qquad f'_{\rm f} = f -  \frac{\left\langle \rho f \right\rangle}{\left\langle \rho \right\rangle}.
\label{eq:Favre triple decomposition definition}
\end{equation}
To simplify the notation and to keep the parallelism with the laminar formulation of
the linearised Navier-Stokes equations, the 
subindex $\left(\cdot\right)_{{\rm f}}$ is dropped hereafter. 

These decompositions are now introduced in the 
conservation equations, Eq.~\eqref{eq:Compressible_NS}, to obtain 
the governing equations for the mean flow (Sec.~\ref{sec:mean flow numerical})
and coherent perturbations (Sec.~\ref{sec:acoustic numerical}).

\subsection{Mean flow}\label{sec:mean flow numerical}
Taking the time average of Eq.~\eqref{eq:Compressible_NS} and 
neglecting terms higher than second-order of the coherent 
component, we obtain the governing 
equations for the mean flow variables, such that
\begin{subequations}
\begin{align}
\nabla \cdot\left(\bar{\rho}\bar{\mathbf{u}}\right)&=0, \\
\nabla \cdot\left(\bar{\rho}\bar{\mathbf{u}} \otimes \bar{\mathbf{u}} \right)&=-\nabla \bar{p} + \nabla\cdot \bar{\greektensor{\uptau}}^{{\rm tot}},\\
\nabla \cdot \left( \bar{\rho} \bar{\mathbf{u}}\bar{e}_t + \bar{\mathbf{u}}\bar{p}\right) &= \nabla\cdot\left(\bar{\mathbf{u}}\cdot\bar{\greektensor{\uptau}}^{{\rm tot}}\right) - \nabla \cdot \bar{\mathbf{q}}^{{\rm tot}},
\end{align}
\label{eq:RANS equations}
\end{subequations}
with 
\begin{equation}
\bar{\greektensor{\uptau}}^{{\rm tot}}=\bar{\greektensor{\uptau}}-\overline{\rho \mathbf{u}' \otimes \mathbf{u}'}
\qquad \text{and} \qquad
\bar{\mathbf{q}}^{{\rm tot}}=\bar{\mathbf{q}}-\overline{\rho \mathbf{u}'c_p T'}.
\label{eq:mean total stress tensor}
\end{equation}
$\bar{\greektensor{\uptau}}$ and $\bar{\mathbf{q}}$ are the  
Favre-averaged stress tensor and heat conduction vector, respectively.
Note that Eq.~\eqref{eq:RANS equations} is not exact and additional 
assumptions are required for its derivation (see~\cite{wilcox1998turbulence} for further details).

Eq.~\eqref{eq:mean total stress tensor} contains two terms that need closure models: 
the Reynolds stress tensor, $-\overline{\rho \mathbf{u}' \otimes \mathbf{u}'}$, and
the turbulent heat flux vector, $-\overline{\rho \mathbf{u}'c_p T'}$.
The Boussinesq approximation~\cite{boussinesq1877essai} is  
used to model the Reynolds stress tensor
by assuming that it is proportional 
to the trace-less mean strain-rate tensor, i. e.
\begin{subequations}
\begin{align}
-\overline{\rho \mathbf{u}' \otimes \mathbf{u}'}= \mu_t \left[ \left(\nabla \bar{\mathbf{u}} 
+ \nabla \bar{\mathbf{u}}^{\top}\right) 
- \frac{2}{3}\left(\nabla \cdot \bar{\mathbf{u}}\right)\greektensor{\updelta}\right]- \frac{2}{3}\bar{\rho} k \greektensor{\updelta},
\end{align}
\end{subequations}
where $\mu_t$ is a scalar termed eddy viscosity and $k$ is the 
turbulent kinetic energy, i.e. $k=\frac{1}{2}\overline{\rho \mathbf{u}'\cdot \mathbf{u}'}$. 

The turbulent heat flux vector is modelled using a Reynolds analogy, such that
\begin{equation}
-\overline{\rho \mathbf{u}'c_p T'}= -\mu_t\frac{c_p}{Pr_t}\nabla \bar{T},
\label{eq:mean turbulent heat flux}
\end{equation}
where $Pr_t$ is the turbulent Prandtl number.

The governing equations are solved using the open-source 
finite volume solver OpenFOAM (version 4.1)~\cite{weller1998tensorial}.
A steady state solution is obtained using the SIMPLE algorithm~\cite{caretto1973two}.
The eddy viscosity, $\mu_t,$ is determined using
the $k$-omega-SST model~\cite{menter2003ten}. 

\subsection{Coherent perturbations}\label{sec:acoustic numerical}
Taking the phase average of Eq.~\eqref{eq:Compressible_NS}, subtracting Eq.~\eqref{eq:RANS equations}, and neglecting terms higher than second-order of the coherent
component, we obtain the governing equations for the coherent perturbations as follows
\begin{subequations}
\begin{align}
&\frac{\partial \tilde{\rho}}{\partial t} + \nabla\cdot\left({\bar{\rho} \tilde{\mathbf{u}}} + \tilde{\rho}\bar{\mathbf{u}}\right) = 0, 
\label{eq:Acoustic mass} \\
&\frac{\partial \bar{\rho} \tilde{\mathbf{u}}}{\partial t}  
+ \left(\bar{\rho}\tilde{\mathbf{u}} + \tilde{\rho}\bar{\mathbf{u}} \right)\cdot \nabla \bar{\mathbf{u}} + \nabla \cdot \left( \bar{\rho} \bar{\mathbf{u}} \otimes \tilde{\mathbf{u}}\right) = -\nabla \tilde{p} + \nabla \cdot\tilde{\greektensor{\uptau}}^{{\rm tot}}, 
\label{eq:Acoustic momentum} \\
&\frac{\partial \tilde{p}}{\partial t} + \nabla \cdot \left( \bar{\mathbf{u}} \tilde{p}+\gamma \tilde{\mathbf{u}}\bar{p}\right)  
+ \left(\gamma-1\right)\left[ \tilde{p} \nabla \cdot \bar{\mathbf{u}} - \tilde{\mathbf{u}} \cdot \nabla \bar{p} \right]
= \left(\gamma-1\right)\left[\tilde{\greektensor{\uptau}}^{{\rm tot}}:\nabla \bar{\mathbf{u}} + \bar{\greektensor{\uptau}}^{{\rm tot}}:\nabla \tilde{\mathbf{u}} - \nabla \cdot \tilde{\mathbf{q}}^{{\rm tot}}\right],
\label{eq:Acoustic energy} 
\end{align}
\label{eq:Acoustic equations}
\end{subequations}
where $:$ is the double dot product, such that $\tensor{M}:\tensor{N}=\mathrm{tr}(\tensor{M}\,\tensor{N}^{\top}).$ The fluctuation total stress tensor is given by
\begin{subequations}
\begin{align}
\tilde{\greektensor{\uptau}}^{{\rm tot}}=\tilde{\greektensor{\uptau}}-\left( \left\langle\rho \mathbf{u}' \otimes \mathbf{u}'\right\rangle - \overline{\rho \mathbf{u}' \otimes \mathbf{u}'}\right),
\end{align}
with
\begin{align}
\tilde{\greektensor{\uptau}} = \mu \left[ \left(\nabla \tilde{\mathbf{u}} 
+ \nabla \tilde{\mathbf{u}}^{\top}\right) 
- \frac{2}{3}\left(\nabla \cdot \tilde{\mathbf{u}}\right)\greektensor{\updelta}\right].
\end{align}
\label{eq:coherent viscous tensor}
\end{subequations}
The fluctuating total heat flux vector is given by 
\begin{subequations}
\begin{align}
\tilde{\mathbf{q}}^{{\rm tot}}=\tilde{\mathbf{q}}-\left(\left\langle \rho \mathbf{u}'c_p T'\right\rangle -\overline{\rho \mathbf{u}'c_p T'}\right),
\end{align}
with
\begin{align}
\tilde{\mathbf{q}}= -\mu\frac{c_p}{Pr}\nabla \tilde{T}.
\end{align}
\end{subequations}
The perturbation temperature is related to pressure and density 
through the linearised equation of state, that reads
\begin{equation}
\tilde{p} = R\left(\bar{\rho}\tilde{T} + \tilde{\rho}\bar{T}\right). 
\end{equation}

These equations (Eq.~\eqref{eq:Acoustic equations}) 
are termed hereafter the coupled formulation 
and are exact (within the assumptions given so far). 
To investigate the effect of mean non-isentropicity on 
the perturbation response of the area expansion, this exact 
formulation (Eq.~\eqref{eq:Acoustic equations}) is further 
simplified in the next section. 

\subsubsection{Uncoupled formulation}\label{sec:Uncoupled formulation}

In this section, we simplify 
the perturbation equations
by neglecting mean-entropy gradients. To this end, the 
energy equation (Eq.~\eqref{eq:Acoustic energy}) is 
given in terms of entropy as
\begin{equation}
\rho T \frac{{\rm D}s}{{\rm D}t} = \greektensor{\uptau}:\nabla \mathbf{u} - \nabla \cdot \mathbf{q}. 
\label{eq:entropy equation}
\end{equation}
For common acoustic applications~\cite{kierkegaard2010frequency,kierkegaard2011simulations,gikadi2012linearized},
the generation of entropy due to viscous effects and thermal conduction can be neglected
in the derivation of the perturbation equations. Additionally, we can assume that entropy fluctuations due to turbulence are negligible, i.e. $s = \bar{s}_{{\rm f}} + \tilde{s}_{{\rm f}}.$
Taking now the phase average $\left\langle \cdot\right\rangle$ of Eq.~\eqref{eq:entropy equation} and subtracting its mean, we obtain, after linearisation, the following equation
\begin{equation}
\frac{{\rm \bar{D}} \tilde{s}}{{\rm D} t} 
+ \left( \frac{\tilde{\rho}}{\bar{\rho}}\bar{\mathbf{u}} + \tilde{u} \right)\cdot \nabla \bar{s} = 0,
\label{eq:entropy lineal}
\end{equation}
where the operator ${\rm \bar{D}}/{\rm D} t=\partial/\partial t + \bar{\mathbf{u}}\cdot \nabla$ denotes the convective derivative associated with the mean flow. Again, the subindex $(\cdot)_{{\rm f}}$ (denoting density-weighted variables) has been dropped to simplify notation. The first term of this equation shows that entropy is convected by the mean flow. The second term represents a source of entropy due to the interaction of acoustic and/or vorticity waves with mean-flow entropy gradients.  

For certain conditions 
(including low-Mach-number 
flows~\cite{kierkegaard2010frequency,kierkegaard2011simulations,gikadi2012linearized}), 
Eq.~\eqref{eq:entropy lineal} can be further simplified by neglecting the mean-entropy
gradient, i.e. $\nabla \bar{s} \approx \mathbf{0},$ leading to a convective equation, such that 
\begin{equation}
\frac{{\rm \bar{D}} \tilde{s}}{{\rm D} t} = 0.
\label{eq:entropy isentropic}
\end{equation}
This result effectively decouples the energy equation from the mass and momentum conservation equations. Eq.~\eqref{eq:entropy isentropic} together 
with Eqs.~\eqref{eq:Acoustic mass} and \eqref{eq:Acoustic momentum} is thus termed 
the uncoupled formulation.

The relation between density, pressure, and entropy can now be obtained by
linearising the Gibbs relation. The resulting expression is then integrated
from a reference state and, subsequently, the phase average of that equation
is taken. Neglecting also temperature fluctuations due to turbulence, i.e. $T'_{{\rm f}}=0,$
we obtain
\begin{equation}
\tilde{p} = \bar{c}^2 \left( \tilde{\rho} + \bar{\rho}\frac{\tilde{s}}{c_p} \right).
\label{eq:Gibbs linearised}
\end{equation} 
This expression is the generalisation of 
previous work~\cite{kierkegaard2011simulations,gikadi2014impact,holmberg2015frequency} 
to include entropy fluctuations. 

For low-Mach-number flows, variations of the mean density $\bar{\rho}$ 
are negligible. Assuming 
also that turbulence fluctuations do not generate significant density 
variations, i.e. $\rho'=0,$ as in previous studies~\cite{gikadi2014impact,holmberg2015frequency},
we can prove that the density-weighted variables given by Eq.~\eqref{eq:Favre triple decomposition definition} simplify to the variables defined in Eq.~\eqref{eq:triple decomposition definition} 
for the standard triple decomposition. This proves that this formalism collapses to
the formulation proposed previously in the 
literature~\cite{gikadi2014impact,holmberg2015frequency} for low-Mach-number flows.

\subsubsection{Closure model}\label{sec:Closure}
Following~\cite{reynolds1972mechanics,alizard2015optimal}, and analogous to 
the closure model used for the mean flow, the fluctuations of the 
Reynolds stress tensor are assumed proportional to the
fluctuations of the strain-rate tensor as
\begin{equation}
-\left(\left\langle\rho \mathbf{u}' \otimes \mathbf{u}'\right\rangle-\overline{\rho \mathbf{u}' \otimes \mathbf{u}'}\right) = \mu_t\left[ \left(\nabla \tilde{\mathbf{u}} 
+ \nabla \tilde{\mathbf{u}}^{\top}\right) 
- \frac{2}{3}\left(\nabla \cdot \tilde{\mathbf{u}}\right)\greektensor{\updelta}\right].
\label{eq:coherent turbulent tensor}
\end{equation}
Additionally, we assume that the turbulence is not affected by the 
acoustic fluctuations which allows the eddy viscosity 
to be considered ``frozen'' and equal to the value obtained for the mean flow.  

The turbulent heat flux vector (or dissipation of internal energy) is modelled
again using a Reynolds analogy, so that
\begin{equation}
-\left( \left\langle \rho \mathbf{u}'c_p T'\right\rangle -\overline{\rho \mathbf{u}'c_p T'}\right)= -\mu_t\frac{c_p}{Pr_t}\nabla \tilde{T}.
\label{eq:total heat flux}
\end{equation}

\subsubsection{Finite element strategy}\label{sec:finite element}
Eqs.~\eqref{eq:Acoustic mass}, \eqref{eq:Acoustic momentum}, and \eqref{eq:Acoustic energy},
are referred as the coupled formulation while Eqs.~\eqref{eq:Acoustic mass}, \eqref{eq:Acoustic momentum}, and~\eqref{eq:entropy isentropic} 
are termed the uncoupled formulation.
The two sets of equations are recast in the frequency domain and solved using a finite
element method stabilised with a least-squares formulation~\cite{donea2003finite}. 
The equations are implemented in the open-source computing framework FEniCS~\cite{alnaes2015a, logg2012a}. After discretisation, the resulting linear system is inverted using the sparse
direct solver MUMPS~\cite{MUMPS:1, MUMPS:2}. The numerical implementation
has been validated with several canonical test cases,
including an isentropic nozzle (see~\ref{sec:Isentropic nozzle})
and the acoustic scattering of an area expansion (\ref{sec:Area expansion acoustic}).

At the inlet and outlet of the domain, non-reflecting boundary 
conditions are imposed. The implementation of the boundary conditions is 
based on the splitting of the flux vector defined at the boundaries 
(that appears naturally in the formulation after integration by parts)
into incoming and outgoing waves in a similar way as that of 
the characteristic method~\cite{poinsot1992boundary}. This method allows
incoming acoustic/entropy waves to be easily imposed at both boundaries while assuring 
non-reflectivity of the outgoing waves. 

\section{Analytical modelling}\label{sec:analytical modelling}
\begin{figure}
\centering
  \begin{tikzpicture}

\useasboundingbox (0.,0.5) rectangle (8.5, 3);


\draw[line width=1] (1.0, 1.0) -- (4.5, 1.0);
\draw[line width=1] (1.0, 2.0) -- (4.5, 2.0);

\node at (1.7, 1.5) {${\rm S}_u$};
\node at (4.9, 2.1) {${\rm S}_b$};


\draw[line width=1] (4.5, 0.5) -- (4.5, 1.0);
\draw[line width=1] (4.5, 2.0) -- (4.5, 2.5);

\draw[line width=1] (4.5, 0.5) -- (9.0, 0.5);
\draw[line width=1] (4.5, 2.5) -- (9.0, 2.5);

\node at (8.3, 1.5) {${\rm S}_d$};
\node at (6.0, 2.8) {${\rm S}_{\rm wall}$};
\node at (3.4, 2.3) {${\rm S}_{\rm wall}$};

\node at (6.0, 1.5) {${\rm V}$};




\draw[line width=0.5, dashed] (2, 1.05) -- (2, 1.95);

\draw[line width=0.5, dashed] (2, 1.05) -- (4.55, 1.05);
\draw[line width=0.5, dashed] (2, 1.95) -- (4.55, 1.95);

\draw[line width=0.5, dashed] (4.55, 0.55) -- (4.55, 1.05);
\draw[line width=0.5, dashed] (4.55, 1.95) -- (4.55, 2.45);

\draw[line width=0.5, dashed] (4.55, 0.55) -- (8.0, 0.55);
\draw[line width=0.5, dashed] (4.55, 2.45) -- (8.0, 2.45);

\draw[line width=0.5, dashed] (8, 0.55) -- (8.0, 2.45);

\end{tikzpicture}
  \caption{Schematic of the region ${\rm V}$ used 
  in the derivation of Eq.~\eqref{eq:conservation model}. ${\rm S}$ is 
  the surface bounding ${\rm V}$: $ {\rm S} = {\rm S}_u \cup {\rm S}_d \cup {\rm S}_b \cup {\rm S}_{\rm wall}$.}
  \label{fig:Domain model}
\end{figure}
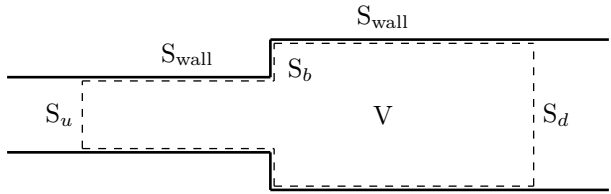
Let us define the control volume ${\rm V}$ bounded by the surface ${\rm S}$
as depicted in Fig.~\ref{fig:Domain model}. The sub-surfaces 
${\rm S}_u$ and ${\rm S}_d$ are taken sufficiently far from the expansion
so that the flow variables can be considered one-dimensional over them.
The conservation equations, given by Eq.~\eqref{eq:Compressible_NS}, are 
integrated over this volume as the starting point for the derivation of the model.
The volume terms are then transformed to surface integrals 
using the divergence theorem.
The flow variables are assumed to evolve quasi-steadily, 
so that $\partial(\cdot)/\partial t = 0.$ This assumption restricts the model 
to low frequencies.
Viscous friction on the wall boundaries (${\rm S}_{{\rm wall}}\; \cup \; {\rm S}_b$)
is neglected. Viscous effects, however, 
are implicitly retained in the rest of the domain. Finally, a non-slip and adiabatic 
boundary condition is imposed on 
the wall boundaries. 
Considering symmetry arguments, the conservation of 
mass, momentum, and energy, can be written~\cite{ronneberger1987}, respectively, as   
\begin{subequations}
\begin{align}
\rho_u u_u A_u &= \rho_d u_d A_d,
\label{eq:conservation model mass}
\\
\left(\rho_u u_u^2 + p_u \right) A_u &= \left( \rho_d u_d^2 + p_d \right) A_d - p_b A_b,
\label{eq:conservation model momentum}
\\
h_{t,u} &= h_{t,d},\label{eq:conservation model energy}
\end{align}
\label{eq:conservation model}
\end{subequations}
where $h_t$ denotes total enthalpy. The relation between total 
energy and enthalpy, $h_t = e_t + p/\rho,$
was used to obtain Eq.~\eqref{eq:conservation model energy}. 

The pressure on the backplate, $p_b,$ in the momentum 
equation (Eq.~\eqref{eq:conservation model momentum}), requires 
closure. A common assumption for separated 
flows~\cite{batchelor1967,cummings1975,ronneberger1987,davies1988} 
is that the pressure acting on the backplate is equal to the pressure just
before the area expansion, i.e. $p_b=p_u.$ 

The decomposition introduced in 
Sec.~\ref{sec:Problem formulation} is now used 
in Eq.~\eqref{eq:conservation model} to obtain the governing 
equations for the mean (Sec.~\ref{sec:mean flow model}) and 
perturbation (Sec.~\ref{sec:acoustic model}) components. 

\subsection{Mean flow}\label{sec:mean flow model}
The governing equations for the mean flow~\cite{gaudron2019acoustic} are
\begin{subequations}
\begin{align}
\beta\Xi M_d &= M_u \eta, \label{eq:Model Mean Mass}\\
\beta\Xi^2\left(1 + \gamma M_d^2 \right) &= 1 + \gamma M_u^2 \eta, \\
\Xi^2\left( 1 + \frac{(\gamma -1)}{2}M_d^2 \right) &= \left( 1 + \frac{(\gamma -1)}{2}M_u^2 \right),
\end{align}
\label{eq:Model Mean flow}
\end{subequations}
where $\beta$ and $\Xi$ are non-dimensional parameters defined as 
$\beta=\bar{\rho}_d/\bar{\rho}_u$ and $\Xi=\bar{c}_d/\bar{c}_u.$
Eqs.~\eqref{eq:Model Mean flow} form a system of three independent, non-linear
equations with six unknowns. This system can be solved numerically 
after imposing three of them. Here, we impose the geometry through $\eta,$
the type of gas through $\gamma,$  and the normalised mass flow rate
through the Mach number at the inlet, $M_u.$ The mean pressure ratio, defined as
$\alpha=\bar{p}_d/\bar{p}_u= \beta \Xi^2,$
could be used alternatively in Eq.~\eqref{eq:Model Mean flow}.

The degree of non-isentropicity of the area expansion 
is described by the mean-entropy
jump across the area expansion, given by
\begin{equation}
\Delta \bar{s}/c_p=\ln\left(\beta^{\frac{(1-\gamma)}{\gamma}}\Xi^{\frac{2}{\gamma}} \right).
\label{eq:Mean entropy jump}
\end{equation}
  

\subsection{Perturbations}\label{sec:acoustic model}
The governing equations for the perturbation part are now 
obtained by linearising Eq.~\eqref{eq:conservation model}. The 
resulting system of equations is rewritten in terms of plane
waves propagating in the same and opposite direction to 
the mean flow, $\hat{p}^+,$ $\hat{p}^-$ and $\hat{s}^+,$
and arranged into input and output vectors to the 
system, written, respectively, as  
$\mathbf{X}=\left[ \hat{p}^+_u/\gamma \bar{p}_u \quad \hat{p}^-_d/\gamma \bar{p}_d \quad \hat{s}^+_u/c_p\right]^{\top},$ and 
$\mathbf{Y}=\left[\hat{p}^+_d/\gamma \bar{p}_d \quad \hat{p}^-_u/\gamma \bar{p}_u \quad \hat{s}^+_d/c_p \right]^{\top}.$ The resulting system of equations is given by
\begin{subequations}
\begin{align}
\mathbf{Y} = \mathbf{A}^{-1}\mathbf{B} \mathbf{X},
\end{align}
\label{eq:simplified linear model}
with
\begin{align}
\mathbf{A}=
\begin{bmatrix}
\beta \Xi \left( M_d + 1 \right)   & \eta\left(1-M_u\right) & - \beta \Xi M_d \\
\beta \Xi^2\left( M_d + 1\right)^2 & \eta\left[1 - \left(1-M_u\right)^2\right] - 1 & - \beta \Xi^2 M_d^2 \\
\beta \Xi^3\left[ \frac{M_d^3}{2} + \frac{3}{2}M_d^2 + \frac{1+\gamma M_d}{(\gamma-1)}\right]  &
\eta\left[ -\frac{M_u^3}{2} + \frac{3}{2}M_u^2 + \frac{1- \gamma M_u}{(\gamma-1)}\right] & 
-\frac{\beta\Xi^3}{2}M_d^3
\end{bmatrix},
\label{eq:A matrix}
\end{align}
and
\begin{align}
\mathbf{B} = 
\begin{bmatrix}
\eta\left(1 + M_u \right) & \beta \Xi\left( 1 - M_d \right) & -\eta M_u \\
\eta\left[\left(1+M_u\right)^2-1\right]+1 & -\beta\Xi^2\left(1-M_d\right)^2 & -\eta M_u^2 \\
\eta\left[\frac{M_u^3}{2} + \frac{3}{2}M_u^2 + \frac{1+\gamma M_u}{(\gamma-1)}\right] &
\beta\Xi^3\left[-\frac{M_d^3}{2} + \frac{3}{2}M_d^2 +\frac{1-\gamma M_d}{(\gamma-1)}\right] &
-\frac{\eta}{2}M_u^3
\end{bmatrix}.
\label{eq:B matrix}
\end{align}
\end{subequations}


For low Mach number flows, i.e. $M_u^2,\; M_d^2 \ll 1,$ the 
solution to Eq.~\eqref{eq:Model Mean flow} is $\Xi=1,$ $\beta=1$ 
and $M_d= M_u \eta.$ At this limit, the acoustic response when an 
entropy wave enters the system is 
identically zero, i.e. $\hat{p}_u^-=0$ and $\hat{p}_d^+=0.$
\section{Mean-flow results}\label{sec:mean flow}

\begin{figure}
\centering
	\subfloat{\begin{tikzpicture}
				\node [inner sep=0pt,above right] 
                {\includegraphics[width=\textwidth]{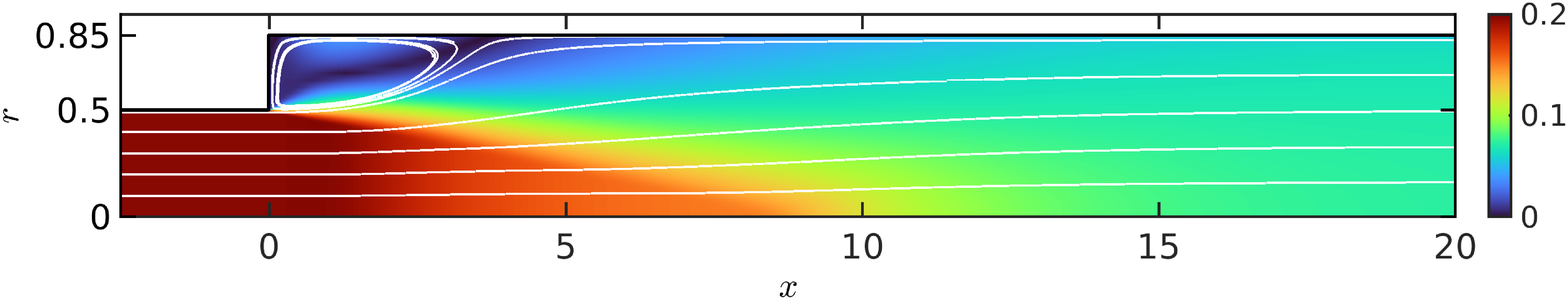}};
                \node at (0,2.6) {(a)};
 			  \end{tikzpicture}} \hspace{0.1mm}
	\subfloat{\begin{tikzpicture}
				\node [inner sep=0pt,above right] 
                {\includegraphics[width=\textwidth]{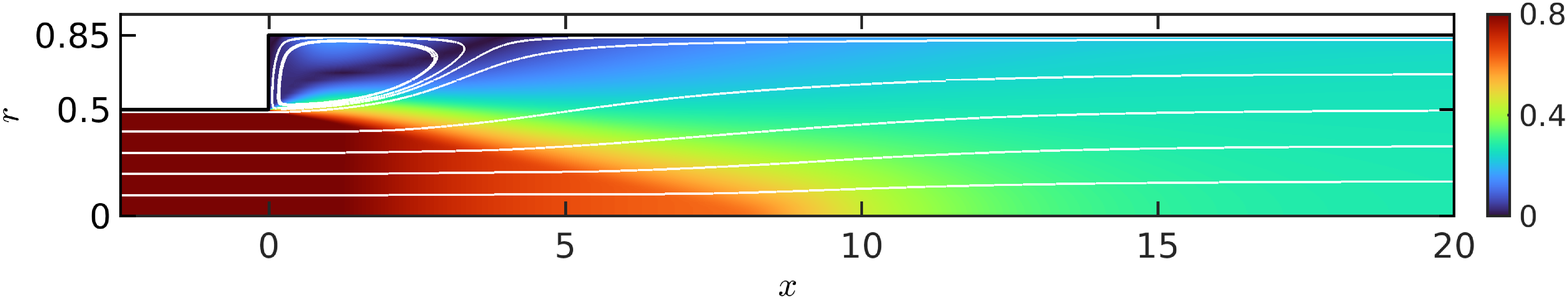}};
                \node at (0,2.6) {(b)};
 			  \end{tikzpicture}} \hspace{0.1mm}
 	\subfloat{\begin{tikzpicture}
				\node [inner sep=0pt,above right] 
                {\includegraphics[width=\textwidth]{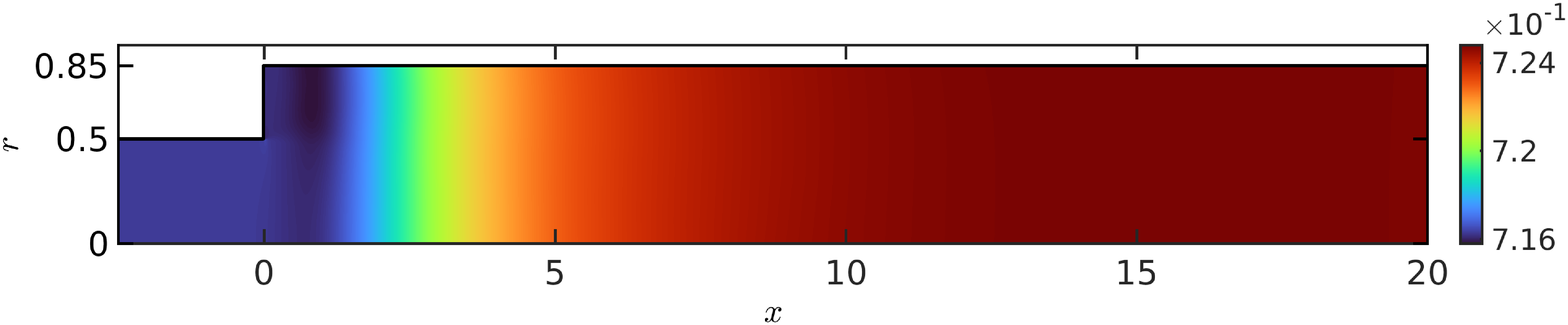}};
                \node at (0,2.6) {(c)};
 			  \end{tikzpicture}} \hspace{0.1mm}
	\subfloat{\begin{tikzpicture}
				\node [inner sep=0pt,above right] 
                {\includegraphics[width=\textwidth]{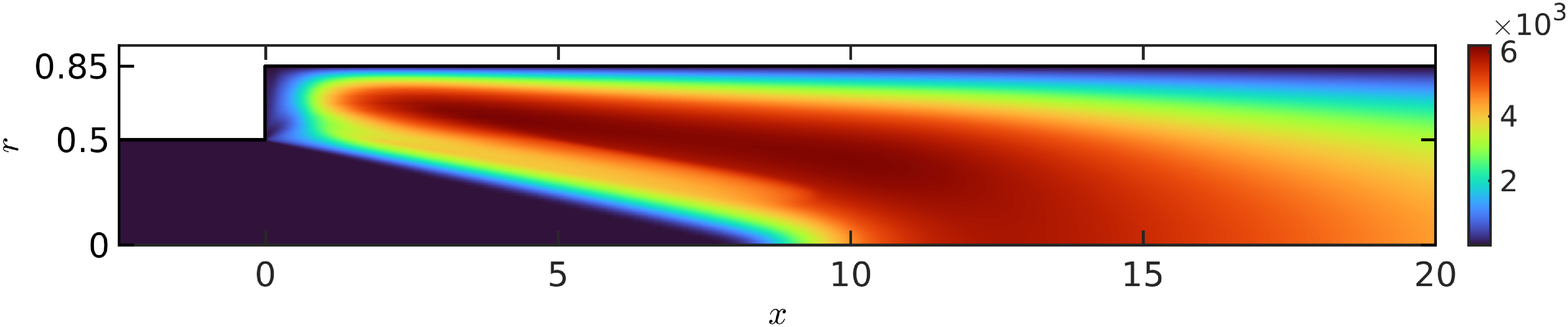}};
                \node at (0,2.6) {(d)};
 			  \end{tikzpicture}} 
\caption{Mean-flow hydrodynamic variables. Normalised norm of the velocity field $|\bar{\mathbf{u}}|/\bar{c}_u,$ 
for (a) $M_u=0.2$ and (b) $M_u=0.8$. White lines denote streamlines. (c) Normalised 
mean pressure, $\bar{p}/\gamma \bar{p}_u,$ and
(d) eddy viscosity, $\mu_t/\mu,$ for $M_u=0.2.$}
\label{fig:Mean flow U}
\end{figure}

In this section, we report the mean-flow results 
obtained for a canonical area expansion as 
introduced in Sec.~\ref{sec:Problem formulation}
using both the numerical and analytical approaches 
outlined in Sec.~\ref{sec:mean flow numerical} 
and Sec.~\ref{sec:mean flow model}, respectively.

The mean-flow variables are determined 
by five non-dimensional parameters: the expansion ratio $\eta,$
the Reynolds number $Re=\left(\bar{\rho}_u\bar{u}_u D_u\right)/\mu,$ the adiabatic index $\gamma,$ 
the Mach number at the inlet $M_u,$ and the Prandtl number $Pr.$

We consider the working fluid to be air, which fixes two of 
the parameters: $\gamma=1.4$ and $Pr=0.7.$
The expansion ratio selected in this paper is $\eta=0.346,$ due to the 
wealth of 
numerical~\cite{kierkegaard2011simulations,gikadi2012linearized,foller2012identification}
and experimental~\cite{ronneberger1987} data available for this geometry. The Reynolds 
number is taken to be $Re=10^6,$ 
which is sufficiently high to assure that the flow topology is independent to this 
parameter (see~\cite{khezzar1986round}). 
Finally, three upstream Mach numbers are investigated
throughout this publication: $M_u=0.2,\;0.5,\; \text{and}\; 0.8.$ 
The range of Mach numbers considered in this study includes subsonic 
high Mach numbers, for which few previous studies exist.
 
The upstream duct is modelled as a laminar, uniform flow. 
To achieve this, we impose
uniform velocity and temperature profiles 
at the inlet, so that the targeted
Mach number, $M_u,$ is obtained. The turbulent kinetic energy 
at the inlet is set to zero, i.e. $k=0.$ A zero-gradient condition is prescribed to 
both the pressure and specific rate of turbulent dissipation (omega).
A slip boundary condition, that reads
$\bar{\mathbf{u}}\cdot \mathbf{n}=0,$ is imposed on the wall of the upstream duct.
A non-slip, adiabatic boundary condition is imposed on the walls at the backplate 
and downstream duct. The viscous sublayer is fully resolved 
on the wall boundaries, which avoids the use of wall functions. At 
the outlet of the domain, uniform pressure is imposed and zero-gradient 
for the rest of variables. 

Due to the axisymmetry of both the domain and boundary conditions, 
the mean flow is assumed axisymmetric. To exploit this, 
the flow is solved in a wedge domain of angle $5 \degree$ with
appropriate boundary conditions in the azimuthal direction. The mesh
extends only one cell in the azimuthal direction. 
The upstream and downstream 
ducts extends $50 D_u$ each in the streamwise direction.
This length was selected to assure that an accurate separation of acoustic waves 
into downstream/upstream-propagating waves was possible at low frequencies. 
All the meshes are structured and composed mostly of hexahedral cells together 
with a layer of prismatic cells in the axis of revolution. A total of 
$276,400$ cells was used for each configuration. The meshes used 
yielded $y^+$ values around $1$ at the duct walls.

Figures~\ref{fig:Mean flow U}(a) and~\ref{fig:Mean flow U}(b) depict the 
velocity field for $M_u=0.2$
and $0.8,$ respectively. The flow remains uniform in the upstream 
duct and separates at the area expansion, creating a large turbulent 
recirculation region at the corner of the domain. For both Mach numbers, the
length of the recirculation zone is $x_R/D_d \approx 2.05,$ which is 
consistent with experimental measurements reported 
by Khezzar et al.~\cite{khezzar1986round}
for a similar geometry. Figures~\ref{fig:Mean flow U}(a) and~\ref{fig:Mean flow U}(b) 
also show that the influence of the Mach number on this length
and the topology of the mean flow is negligible. 

In the central 
part of the domain, a jet is formed. This jet remains almost parallel 
just after the end of the upstream duct. This translates to the
pressure just before the expansion being equal to the pressure acting 
on the backplate as shown in figure~\ref{fig:Mean flow U}(c).
The central jet is separated from the 
low-speed recirculation zone by an expanding shear layer. At the end of 
the recirculation region, this shear layer
merges with the boundary layer on the duct wall.
After that point, 
the velocities at the centre and periphery of the duct slowly converge
forming a fully-developed turbulent pipe flow at about $20 D_u$ downstream of
the expansion. This whole process creates strong turbulence as visualised in the form
of eddy viscosity in figure~\ref{fig:Mean flow U}(d).

\begin{figure}
\centering
	\subfloat{\begin{tikzpicture}
				\node [inner sep=0pt,above right] 
                {\includegraphics[width=\textwidth]{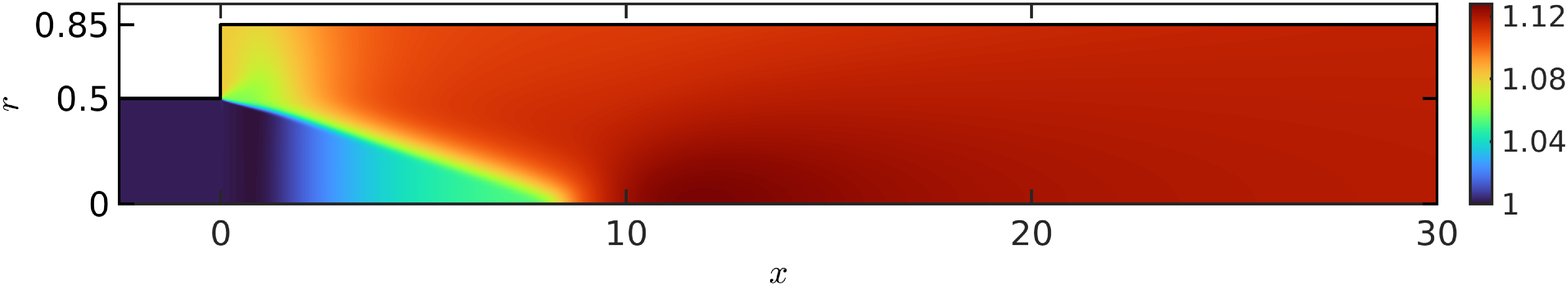}};
                \node at (0,2.6) {(a)};
 			  \end{tikzpicture}} \hspace{0.1mm}
	\subfloat{\begin{tikzpicture}
				\node [inner sep=0pt,above right] 
                {\includegraphics[width=\textwidth]{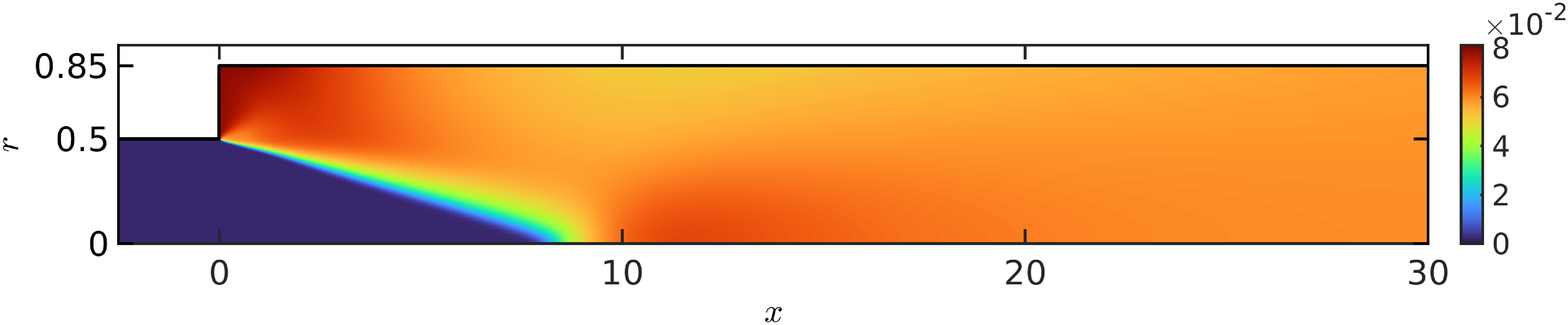}};
                \node at (0,2.6) {(b)};
 			  \end{tikzpicture}}
\caption{Mean-flow thermodynamic variables. (a) Normalised mean-flow 
temperature, $\bar{T}/\bar{T}_u,$ 
and (b) mean entropy increment, $\Delta\bar{s}/c_p,$ for $M_u=0.8.$}
\label{fig:Mean flow Entropy}
\end{figure}

The strong turbulent motion across the shear layer dissipates kinetic energy into
heat by friction effects. This leads to a sharp increase of temperature and
entropy across it as depicted in figure~\ref{fig:Mean flow Entropy}.
The variation of entropy, which is almost 
constant in the upstream and downstream ducts, occurs in a very thin region 
and can be thought as a jump. This region will act
as a source of entropy perturbations 
in the presence of acoustic/vortical waves, as 
described by Eq.~\eqref{eq:entropy lineal}.
\begin{figure}
\centering
	\subfloat{\begin{tikzpicture}
				\node [inner sep=0pt,above right] 
                {\includegraphics[width=\textwidth]{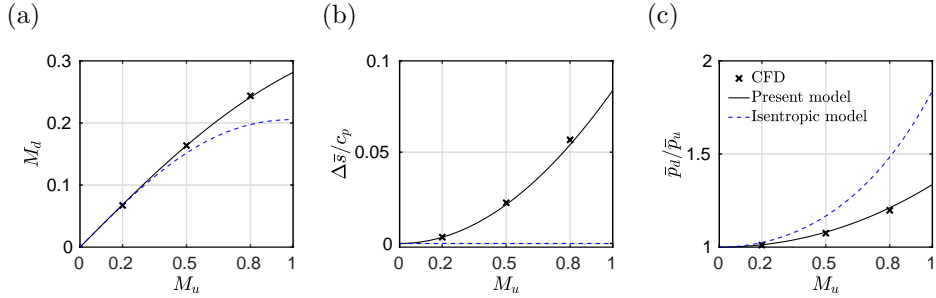}};
              \node at (0,3.7) {(a)};
			  \node at (4.2,3.7) {(b)};
              \node at (8.5,3.7) {(c)};
 			  \end{tikzpicture}} 
\caption{Model performance: (black crosses) 
computational results (CFD), (black solid line) present analytical
model, and (blue dashed line) isentropic model. (a) Downstream Mach 
number, $M_d,$ (b) entropy jump , $\Delta \bar{s}/c_p,$ 
and (c) mean pressure ratio, $\bar{p}_d/\bar{p}_u.$ }
\label{fig:Mean flow Model}
\end{figure}

The performance of the model is now assessed against
the numerical results. To this end, the flow 
variables in the upstream and downstream ducts are 
sampled at the cross-sectional 
locations $x=-1$ and $x=30$, respectively, 
and averaged over the surfaces. In the upstream duct, 
the flow variables are visually one-dimensional 
right up until the expansion and variations 
in the axial direction are negligible as observed 
in figures~\ref{fig:Mean flow U} and~\ref{fig:Mean flow Entropy}.
This justifies the selection of the sampling location very 
close to the area expansion. Figure~\ref{fig:Mean flow Model} shows that 
the agreement between the numerical results and the 
model predictions is excellent for all the range of subsonic 
Mach numbers. Figure~\ref{fig:Mean flow Model} also shows that, 
at low Mach numbers, the strength of the entropy
jump is low and, thus, the values of the 
mean-flow variables are close to the 
predictions for an insentropic expansion.
In contrast, when the Mach number increases, the strength of 
the entropy jump rapidly increases and the results 
deviate from isentropic theory.
The increase of entropy is in fact directly linked to losses of
stagnation pressure. This is observed in 
figure~\ref{fig:Mean flow Model}(c) that shows that, at high 
Mach numbers, the pressure recovered
along the recirculation region is significantly lower for the 
sudden area expansion than for an isentropic expansion.

\section{Scattering of entropy waves}\label{sec:entropy}
\begin{figure}
\centering
	\subfloat{\begin{tikzpicture}
				\node [inner sep=0pt,above right] 
                {\includegraphics[width=\textwidth]{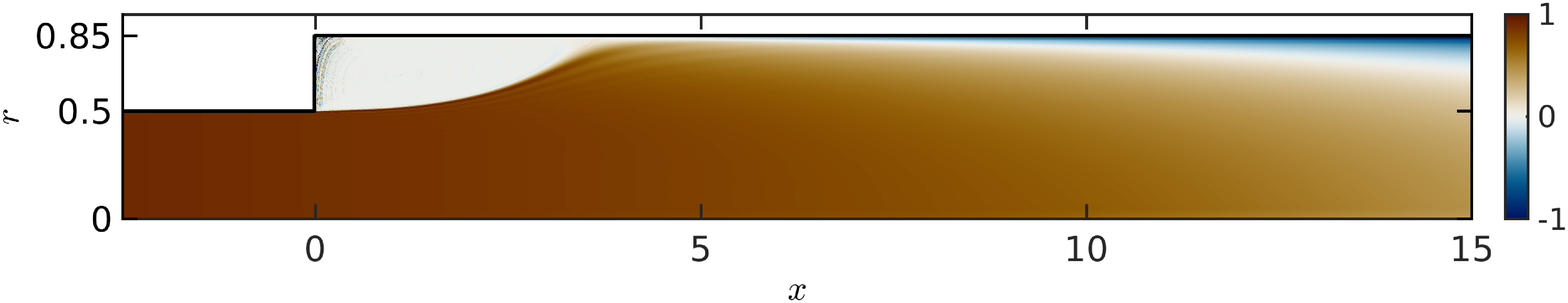}};
                \node at (0,2.6) {(a)};
 			  \end{tikzpicture}} \hspace{0.1mm}
	\subfloat{\begin{tikzpicture}
				\node [inner sep=0pt,above right] 
                {\includegraphics[width=\textwidth]{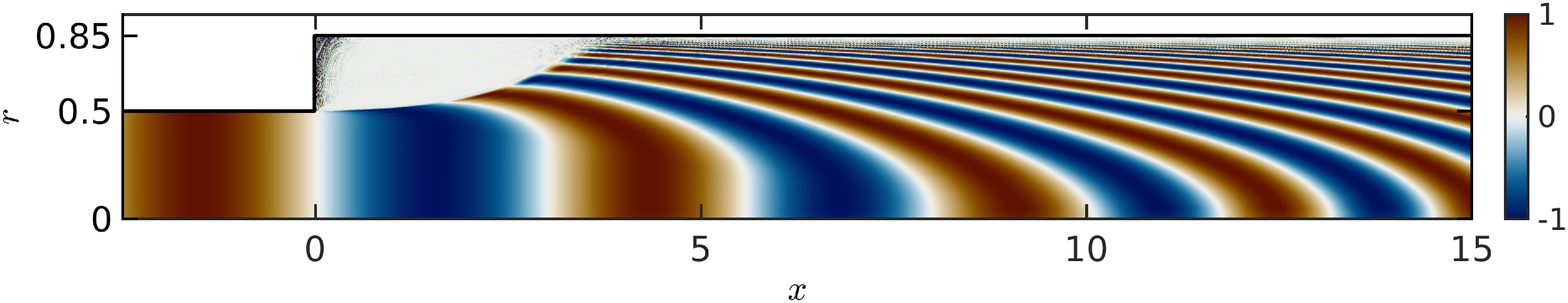}};
                \node at (0,2.6) {(b)};
 			  \end{tikzpicture}} \hspace{0.1mm}
    \subfloat{\begin{tikzpicture} 
				\node [inner sep=0pt,above right] 
                {\includegraphics[width=\textwidth]{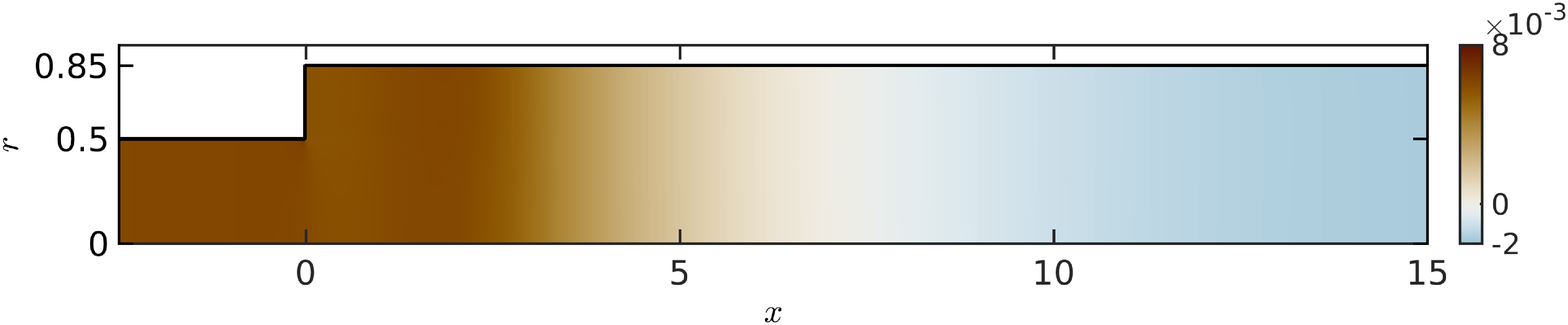}};
                \node at (0,2.6) {(c)};
 			  \end{tikzpicture}} \hspace{0.1mm}
\caption{Real part of the entropy perturbation for: (a) $St=0.025$ and (b) $St=1$. 
(c) Real part of the perturbation pressure at $St=0.025$ and $M_u=0.2.$ The 
results were obtained with the coupled solver.}
\label{fig:Entropy Distribution}
\end{figure}
In this section, the acoustic response of the expansion to incoming 
entropy waves is investigated using the numerical and analytical 
approaches introduced in Sec.~\ref{sec:acoustic numerical} and 
Sec.~\ref{sec:acoustic model}, respectively. 

We first characterise the expansion numerically, for which the mean-flow 
variables presented in Sec.~\ref{sec:mean flow} are interpolated 
to an acoustic mesh. This mesh is two-dimensional, fully 
unstructured and composed of $558,339$ triangles. The upstream
and downstream ducts
are $15 D_u$ and $40 D_u$ long, respectively. The coupled and decoupled formulations
(see Sec.~\ref{sec:finite element}) of the linearised equations are 
formulated in cylindrical coordinates and solved assuming axisymmetry 
of the flow variables. The numerical implementation allows for the 
order of the basis functions to be selected from first- to fifth-order 
Lagrangian polynomials. The results presented here were obtained using
second-order polynomials.     

An incoming entropy perturbation of 
normalised amplitude $\sigma=\hat{s}^+_u/c_p$
is imposed at the inlet of the domain. Non-reflecting 
boundary conditions are prescribed at both ends,
so only reflected and transmitted plane acoustic 
waves exist at the far upstream and downstream ducts, respectively.
At the backplate, a rigid wall 
non-slip boundary condition is imposed, namely $\tilde{\mathbf{u}}=\mathbf{0}$. 
Following~\cite{kierkegaard2011simulations}, 
a slip boundary condition, that reads $\tilde{\mathbf{u}}\cdot \mathbf{n}=0,$ 
is prescribed on the walls of the upstream and downstream
ducts. This approach reduces the computational cost as 
the acoustic boundary layer does not need to be 
resolved. However, any viscous dissipation 
associated with the walls is not captured. Gikadi et al.~\cite{gikadi2012linearized} 
showed that, for this geometry, the acoustic results were
the same using either slip or non-slip boundary conditions on the duct walls.

We consider frequencies ranging $8 \times 10^{-4} \leq He \leq 1.2$ for 
$M_u=0.2,$ and $8 \times 10^{-4} \leq He \leq 3.0$ for $M_u=0.5$ 
and $M_u=0.8.$ In the upstream duct, the shortest acoustic 
wavelength is $\lambda \approx 0.42$ obtained for $M_u=0.8$ 
and $He=3.0$ and for which the resolution is approximately $10$ elements 
per wavelength. In the downstream duct, the 
shortest acoustic wavelength 
is $\lambda \approx 2.5$ obtained for $M_u=0.5$ 
and $He=3.0,$ which yields a resolution of 
around $63$ elements per wavelength at 
the coarsest section of the duct. The 
shortest wavelength for the entropy at the upstream duct 
is $\lambda \approx 1.04$ obtained for $He=3.0$
and $M_u=0.5.$ There are approximately $21$ elements per 
entropy wavelength at this frequency. A detailed analysis of 
the error obtained for a given number of points per 
wavelength and polynomial degree for the current implementation
can be found in~\ref{sec:numerical properties}.

Figures~\ref{fig:Entropy Distribution}(a) 
and~\ref{fig:Entropy Distribution}(b)
show the spatial distribution of entropy obtained using
the coupled formulation of the linearised Navier-Stokes
equations (LNSE). Neglecting viscous dissipation and 
heat conduction, the evolution of the perturbation entropy is
given by Eq.~\eqref{eq:entropy lineal}. This equation
comprises two terms: a convective term, that describes the 
advection of entropy by the mean flow as a passive scalar, and 
a source term that describes the generation of perturbation 
entropy through the interaction of acoustic/vorticity waves 
with mean-entropy gradients. At low Mach numbers, 
figure~\ref{fig:Mean flow Model}(b) shows that the mean-entropy 
jump for this configuration is weak and, thus, the source term
in the aforementioned equation is negligible. In this case, 
the perturbation entropy is simply advected by the mean flow as
observed in figures~\ref{fig:Entropy Distribution}(a) 
and~\ref{fig:Entropy Distribution}(b) 
for two different frequencies. For both frequencies, the
entropy presence expands after the expansion, but does so gradually,
meaning that there is an annular region in the recirculation zone 
where the entropy is negligible. At higher Mach numbers, the source
term becomes significant and 
further perturbation entropy is generated after the 
expansion in the thin region where the mean entropy 
sharply increases (see figure~\ref{fig:Mean flow Entropy}(b)).  

Yang et al.~\cite{yang2020entropy}
showed that, at low frequencies and low Mach numbers, 
the main acoustic source term 
due to entropy is a dipole given by 
\begin{equation}
\text{Source} = \frac{1}{\bar{\rho}} \nabla \cdot \left( \nabla \bar{p} \frac{\tilde{s}}{c_p}\right).
\label{eq:source}
\end{equation}
The mean pressure, $\bar{p}$ is constant upstream of the area expansion, 
it then recovers along the recirculation region -- in the streamwise direction --
and becomes constant again after it, as shown 
in figure~\ref{fig:Mean flow U} (c). The entropy perturbation, $\tilde{s}$, and, hence, the entropy-related 
source term (Eq.~\eqref{eq:source}) 
are only significant within the expanding jet, with negligible
presence in the recirculation zone. Figure~\ref{fig:Entropy Distribution}(c) 
shows an example of acoustic field generated by the entropy fluctuations.

To compute the reflection, $P^-_u/\sigma$, and transmission, $P^+_d/\sigma$,
coefficients, the magnitude of the acoustic plane waves is extracted
using the multi-microphone method~\cite{seybert1977experimental,poinsot1986experimental}.
To this end, two post-processing zones are selected: $-15 \leq x \leq-5$ 
upstream and $25 \leq x \leq 38$ downstream of the area expansion. The 
variables are averaged over the cross-sectional area at each location
to reduce the influence of numerical disturbances. $201$ and $131$ 
equispaced points 
are used in the upstream and downstream ducts, respectively.

\begin{figure}
\centering
	\subfloat{\begin{tikzpicture}
				\node [inner sep=0pt,above right] 
                {\includegraphics[width=\textwidth]{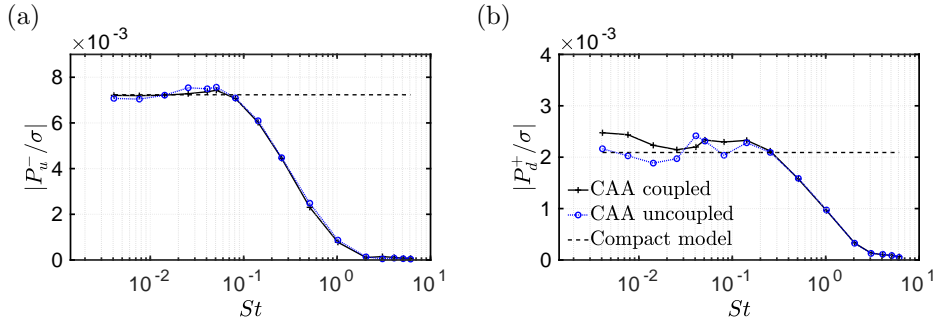}};
                \node at (0, 4.0) {(a)};
                \node at (6.25,4.0) {(b)};
 			  \end{tikzpicture}} 
\caption{Acoustic (a) reflection, $P^-_u/\sigma,$ and (b) transmission coefficients, $P^+_d/\sigma,$ for $M_u=0.2.$ Numerical results using (black plus signs) coupled and (blue circles) uncoupled formulations, and (black dashed line) analytical model.
}
\label{fig:Entropy Mu 0.2}
\end{figure}

\begin{figure}
\centering
	\subfloat{\begin{tikzpicture}
				\node [inner sep=0pt,above right] 
                {\includegraphics[width=\textwidth]{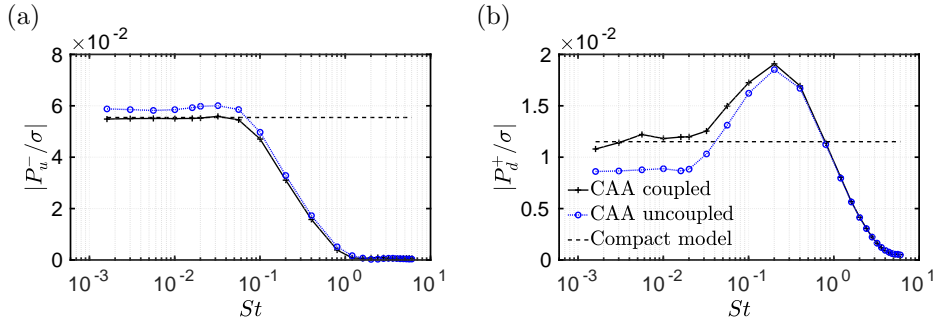}};
                \node at (0, 4.0) {(a)};
                \node at (6.25,4.0) {(b)};
 			  \end{tikzpicture}} 
\caption{Acoustic (a) reflection, $P^-_u/\sigma,$ and (b) transmission 
coefficients, $P^+_d/\sigma,$ for $M_u=0.5.$ Numerical results using (black plus signs) coupled and (blue circles) uncoupled formulations, and (black dashed line) analytical model.
}
\label{fig:Entropy Mu 0.5}
\end{figure}

\begin{figure}
\centering
	\subfloat{\begin{tikzpicture}
				\node [inner sep=0pt,above right] 
                {\includegraphics[width=\textwidth]{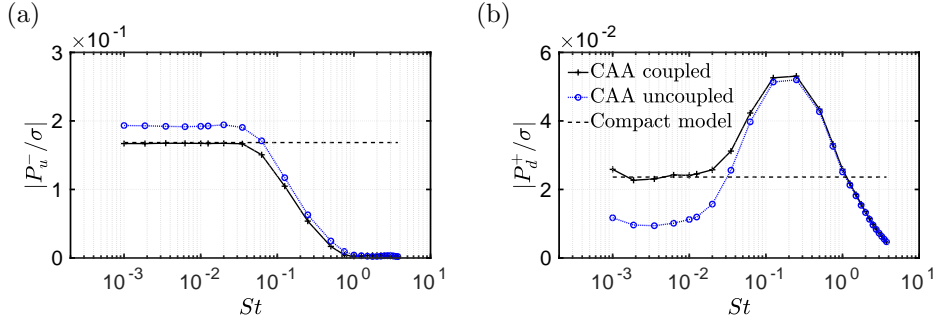}};
                \node at (0, 4.0) {(a)};
                \node at (6.25,4.0) {(b)};
 			  \end{tikzpicture}} 
\caption{Acoustic (a) reflection, $P^-_u/\sigma,$ and (b) transmission coefficients, $P^+_d/\sigma,$ for $M_u=0.8.$ Numerical results using (black plus signs) coupled and (blue circles) uncoupled formulations, and (black dashed line) analytical model.
}
\label{fig:Entropy Mu 0.8}
\end{figure}

Figures~\ref{fig:Entropy Mu 0.2},~\ref{fig:Entropy Mu 0.5}, 
and~\ref{fig:Entropy Mu 0.8} 
show the magnitude of the reflection and transmission coefficients
for $M_u=0.2,$ $M_u=0.5,$ and $M_u=0.8,$ respectively. The reflection 
coefficient behaves as a low-pass filter: at low frequencies, the 
reflection amplitude is nearly constant and, when the frequency increases
non-compact effects cause a drop of its value. The drop-off can be
explained by Eq.~\eqref{eq:source} and 
figures~\ref{fig:Entropy Distribution}(a) and~\ref{fig:Entropy Distribution}(b): 
when more than a half period of the entropy wave is contained 
within the source region, cancelling effects between positive 
and negative parts of the wave occur. As for the transmission 
coefficient, its amplitude remains
nearly constant at low frequencies, then a bump appears at moderate frequencies
to finally drop at higher frequencies. The ratio between the maximum value
at the peak and the value at low frequencies increases with increasing 
Mach numbers. For high Mach numbers, the value of the peak almost 
doubles the value at compact frequencies.
  
At low frequencies, the agreement of the model predictions with 
the numerical results (coupled formulation) is excellent 
for the three Mach numbers. Note that the coupled equations are 
an exact formulation (within the general 
assumptions given in Sec.~\ref{sec:acoustic numerical}) of the problem
and the results obtained with it should be used as reference. The 
uncoupled formulation, on the other hand, neglects the 
mean-entropy gradient and its predictions should be used to quantify 
the validity of that assumption. For 
$M_u=0.2,$ the results obtained for the coupled and uncoupled 
solvers are very close. Some discrepancies between the formulations
are observed for the transmission coefficient, that can be attributed 
to post-processing errors in the computation of the 
magnitude of the coefficients. This error is produced by a combination
of: (i) the very long acoustic wavelengths that turns the application
of the multi-microphone method into an ill-posed problem and (ii) the
very low values of the transmission coefficient and, consequently, the low 
ratio between the actual signal and noise, either numerical (introduced by 
instance by the interpolation of the mean flow) or vortical waves. 
When the Mach number increases, the results of the coupled and decoupled solvers
show some discrepancies. The reason is that
the decoupled formulation neglects the perturbation 
entropy source related to the 
mean-entropy gradient and, as observed in figure~\ref{fig:Mean flow Model}(b),
this becomes significant when increasing the Mach number.

\begin{figure}
\centering
	\subfloat{\begin{tikzpicture}
				\node [inner sep=0pt,above right] 
                {\includegraphics[width=\textwidth]{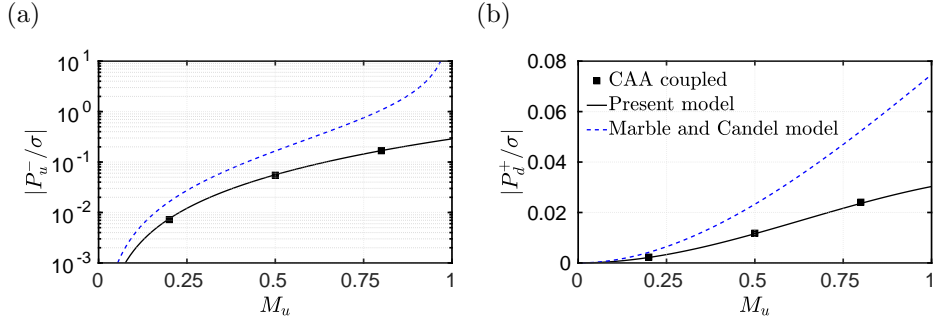}};
                \node at (0, 4.0) {(a)};
                \node at (6.25,4.0) {(b)};
 			  \end{tikzpicture}} \hspace{01mm}
\caption{Acoustic (a) reflection, $P^-_u/\sigma,$ 
and (b) transmission coefficients, $P^+_d/\sigma,$ predicted
by (black solid line) the compact present model, (blue dashed line) the 
Marble and Candel model~\cite{marble77}, and 
(black squares) numerical results at $St=10^{-2}.$}
\label{fig:Model Entropy Comparison}
\end{figure}

In figure~\ref{fig:Model Entropy Comparison}, the numerical results obtained
at low frequencies are compared to the predictions of
the compact model derived by Marble and Candel~\cite{marble77}
for ducts with smooth 
area variations, i.e. nozzle flows. Because this model is compact, 
its predictions only depend on the conditions
upstream and downstream of the area 
expansion. If the assumptions used for its derivation are 
valid, it should be a suitable candidate for 
sudden area expansions as considered here. These assumptions are
that the flow is inviscid, 
quasi-one-dimensional, and isentropic. 
For its derivation, Marbel and Candel imposed 
the conservation of mass, stagnation enthalpy, and entropy. The value 
of the reflection coefficient predicted by that model is higher than 
the actual value across the whole range of Mach numbers. The mismatch 
is significant at low Mach numbers (for $M_u=0.2$ the relative error is 
approximately $117\%$) and it 
becomes very severe when the Mach number is large: Marble and Candel predictions
are two orders of magnitude higher that the predictions of the proposed model
when the Mach number is close to unity. The predictions of the 
transmission coefficient also differ significantly, but the values remain
of the same order of magnitude. The mismatch slightly increases with increasing 
Mach number: the relative error for $M_u=0.2$ is approximately $94\%$ and
increases to $121\%$ for $M_u=0.8$. 

The strong discrepancy between the predictions of the
quasi-1D, isentropic model of Marble
and Candel and the actual computational results was explained by 
Yang et al.~\cite{yang2020entropy} for low-Mach-number area expansions.
They showed that the entropy-related acoustic response
has a very strong dependence on the length of the recirculation 
region: when the flow reattaches 
immediately after the separation, i.e. the 
geometrical expansion matches the flow expansion, 
the values of the acoustic coefficients tend to the predictions of the
Marble and Candel model. However, they significantly deviate when 
the length of the recirculation zone is increased. There 
exists a length after which an increase of such length does not 
bring any changes to the values of the acoustic coefficients.
In terms of acoustic sources, the previous observation states  
that the Marble and Candel 
model is valid when the source is located at the actual expansion, 
but not when the entropy-related source is moved to a different position, 
in this case, into the downstream duct.      

\begin{figure}
\centering
	\subfloat{\begin{tikzpicture}
				\node [inner sep=0pt,above right] 
                {\includegraphics[width=\textwidth]{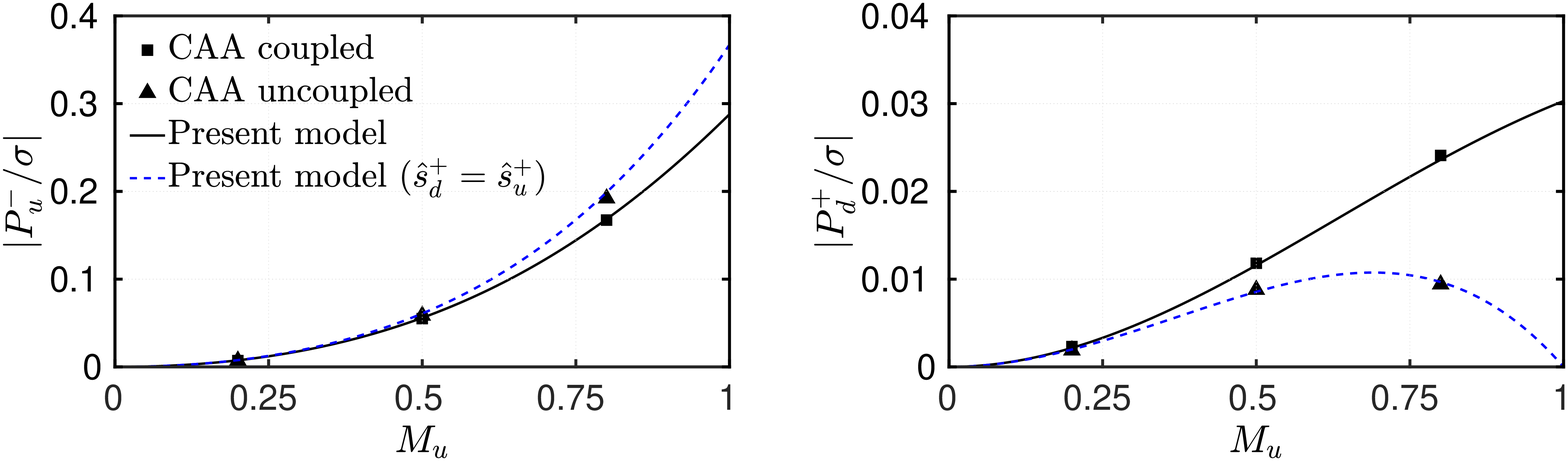}};
                \node at (0, 4.0) {(a)};
                \node at (6.25,4.0) {(b)};
 			  \end{tikzpicture}} 
\caption{Acoustic (a) reflection, $P^-_u/\sigma,$ 
and (b) transmission coefficients, $P^+_d/\sigma,$ predicted
by (black solid line) the compact present model, (blue dashed line) the 
same model neglecting perturbation entropy sources, and 
numerical results at low frequencies using the (squares) coupled 
and (triangles) uncoupled solvers.}
\label{fig:Model Entropy Comparison Isentropic}
\end{figure}

The misalignment of the geometry and the acoustic 
source is sufficient to explain the discrepancies observed at 
low Mach numbers. However, when the Mach number increases, 
the mean-entropy gradient increases, and, thus, there exists a source
of fluctuation entropy within the domain (see Eq.~\eqref{eq:entropy lineal}). 
In other words, the isentropicity assumption used 
in the Marble and Candel model is no longer valid. To explore the 
importance of this assumption, we slightly modify the model
proposed in Sec.~\ref{sec:analytical modelling}. The model relies on 
the mean flow variables obtained using the equations 
presented in Sec.~\ref{sec:mean flow model}. The 
perturbation variables are obtained 
by imposing the linearised conservation 
of mass and momentum 
(Eqs.~\eqref{eq:conservation model mass} 
and~\eqref{eq:conservation model momentum}), but instead 
of the conservation of stagnation enthalpy, we impose the conservation of
perturbation entropy, that reads $\tilde{s}_u = \tilde{s}_d.$ This model 
captures the source/geometry mismatch through the momentum equation
and the hypothesis $\tilde{p}_u = \tilde{p}_d,$ but also imposes the linear 
isentropicity condition. Figure~\ref{fig:Model Entropy Comparison Isentropic} 
shows the predictions of the model. We observe that the  
model predicts values close to the results obtained 
at low frequencies with the uncoupled numerical 
solver. This is expected, since both approaches 
neglect perturbation entropy sources. The values of both  
the reflection and transmission coefficients at low and 
intermediate Mach numbers obtained when neglecting the 
sources of perturbation entropy are close to the actual 
values (the results of the coupled solver). When the Mach number 
goes beyond $M_u\sim0.5,$ the values diverge. The disagreement is especially
large for the transmission coefficient: the actual value monotonically 
increases until unitary Mach, while the isentropic model drops 
after $M_u\approx0.7$ reaching a zero value when the Mach number
reaches one. As for the reflection coefficient, the constant-entropy model 
slightly overpredicts the actual value (a relative error of about $18\%$ is 
obtained for $M_u=0.8$), but the general trend is correctly predicted. These
results show that at low and moderate Mach numbers, the geometry/source 
misalignment is the dominant factor to explain the strong divergence between
the Marble and Candel model and the numerical predictions obtained for 
area expansions. At high Mach numbers, non-isentropic effects become 
significant: for the reflection coefficient, perturbation entropy sources
are still a second order effect, but they highly 
impact the transmission coefficient. 

\begin{figure}
\centering
	\subfloat{\begin{tikzpicture}
				\node [inner sep=0pt,above right] 
                {\includegraphics[width=\textwidth]{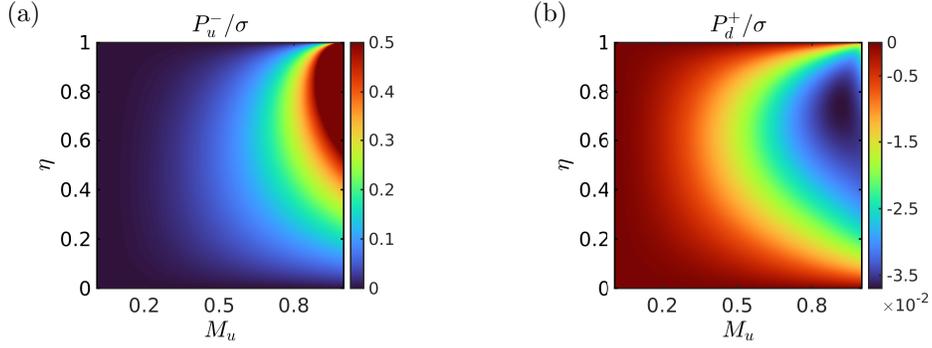}};
                \node at (0, 4.5) {(a)};
                \node at (7.0,4.5) {(b)};
 			  \end{tikzpicture}} 
\caption{Acoustic (a) reflection, $P^-_u/\sigma,$ 
and (b) transmission coefficients, $P^+_d/\sigma,$ predicted
by the present model. The colormap for the reflection coefficient is 
saturated at $P^-_u/\sigma=0.5$: the maximum value 
corresponds to $P^-_u/\sigma = 8.2641$ obtained for $M_u=0.999$ and $\eta = 0.998.$
}
\label{fig:Entropy Model Parametric}
\end{figure}

Finally, we present a 
parametric study performed with the model presented 
in Sec.~\ref{sec:analytical modelling}. Figure~\ref{fig:Entropy Model Parametric}(a),
shows that the reflection coefficient increases with the Mach number and the
expansion ratio, with the maximum obtained for $M_u \to 1$ and $\eta \to 1.$ This 
trend is in clear contrast to the predictions of the Marble and Candel model that 
predicts the maximum for large area expansions, namely $\eta \to 0.$ The value of the 
absolute maxima is also very different, for instance Marble and Candel predicts a 
maxima of $P^-_u/\sigma = 498.3$ for $M_u=0.999$ and $\eta=0.001,$ while 
figure~\ref{fig:Entropy Model Parametric}(a) shows a maximum value of $P^-_u/\sigma = 8.2641$ obtained for $M_u=0.999$ and $\eta = 0.998.$ The transmission coefficient is showed in figure~\ref{fig:Entropy Model Parametric}(b). 
A maximum value of $|P^+_d/\sigma|\approx 0.037$ is obtained at $M_u=0.92$ 
and $\eta=0.73.$ Again, this is different from the Marble and Candel model that
predicts a monotonically increase of the transmission coefficient with the Mach number.  

\section{Conclusions}\label{sec:conclusion}

The acoustic response of a generic area expansion 
subjected to incoming entropy waves was 
investigated both numerically and analytically.
The numerical approach is based on a triple 
decomposition of the flow variables into 
a steady mean, a coherent perturbation part (containing 
acoustic, vortical, and entropy waves),
and a stochastic turbulent part. Separated governing 
equations for each part were derived. The mean
flow was obtained as the solution of the 
Reynolds-Averaged Navier-Stokes (RANS) equations closed
by the $k$-omega-SST turbulence model. The amplitude of the 
coherent perturbations was assumed to be small and the 
equations governing this part were linearised and solved in the 
frequency domain. To account for the effect of turbulence 
on the coherent perturbation, a frozen eddy viscosity  model was used. 
From a post-processing point of view, the use of the 
eddy viscosity eases the computation of the acoustic waves
because it damps the vortical waves generated at the expansion. 
Two different formulations of the perturbation equations were adopted 
in this work: (i) an exact formulation, that was termed a coupled solver,
and (ii) a formulation that decouples the energy equation from 
the mass and momentum conservation equations by 
neglecting mean-entropy gradients, termed an uncoupled solver.

The results show that the area expansion behaves as a low pass-filter 
to entropy fluctuations. The reflection coefficient has a virtually
constant value at low frequencies and drops at higher frequencies 
due to non-compact effects. The transmission coefficient, on the other hand,
also has a constant value at low frequencies, but increases its value at 
intermediate frequencies before dropping at higher frequencies. This 
behaviour was found to be consistent across the range of subsonic 
Mach numbers studied here. The values of the coefficients at 
low frequencies were found to be significantly over-predicted by the  
quasi-one-dimensional and isentropic theory of 
Marble and Candel~\cite{marble77}. The mismatch is especially pronounced
for the reflection coefficient and for subsonic high Mach numbers (for $M_u=0.8$ 
the predictions of the model are one order of magnitude higher than the 
actual values). 

Two main physical mechanisms can be responsible for the disagreement: 
(i) the mean-flow non-isentropicity 
(in other words, stagnation pressure losses)~\cite{de2019generalised} or the 
spatial misalignment between the source term and the geometrical area expansion
owe to the presence of the recirculation zone~\cite{yang2020entropy}. 
To explore the relative importance of each of them, numerical simulations were 
performed with the decoupled solver. This solver neglects the 
mean-entropy gradient term in the energy and, hence, any source of 
perturbation entropy. For the reflection coefficient, it was found that,
for low to moderate Mach numbers, the predictions of the coupled 
and uncoupled solvers were similar. Some disagreement was observed for
higher Mach numbers ($M_u \gtrsim 0.7$), but not enough 
to explain the discrepancies with Marble and Candel. 
These results suggest that this model fails to predict the reflection 
coefficient of area expansions mainly owe to the 
spatial delay of the source term associated with separation. 
For the transmission coefficient,
on the other hand, the trends predicted for both solvers are similar 
up to $M_u\approx0.6,$ where they diverge. After that point, the 
errors of the Marble and Candel model and the uncoupled solver are 
similar. This suggests, that for low to moderate Mach numbers, the 
main factor of discrepancy is again the spatial delay associated 
with separation, but beyond that point the mean flow non-isentropicity
is as important.

An alternative compact model was developed based on the 
conservation of mass, momentum, and stagnation enthalpy. The model 
captures the three-dimensional effects through the 
momentum equation. The agreement between the numerical results and the 
model predictions is excellent for all the range of subsonic Mach numbers studied.
Finally, a parametric study of the acoustic-entropic response 
of the area expansion was performed based on the model. It was found
that the reflection coefficient for large area expansions ($\eta\lesssim 0.2$)
tends to zero for any subsonic Mach number. This is in clear opposition to 
divergent nozzles with large area expansions, where this coefficient monotonically increases with 
decreasing expansion ratios. This result has important implications 
when modelling entropy acceleration through orifices using two-step
models for perforations as showed by 
De Domenico et al.~\cite{de2019generalised,de2021compositional}



\section*{Acknowledgements}
The authors would like to gratefully acknowledge the 
European Research Council (ERC) Consolidator Grant 
AFIRMATIVE (2018-2023) and the Engineering 
and Physical Sciences Research Council (EPSRC) Grant 
CHAMBER (2017-2020) for supporting the current research.

\appendix
\section{Validation case: Isentropic nozzle}\label{sec:Isentropic nozzle}
\begin{figure}
\centering
\includegraphics[width=\textwidth]{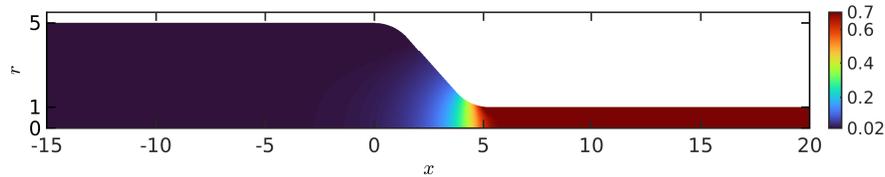}
\caption{Local Mach number $M=\bar{u}/\bar{c}$.}
\label{fig:Mean flow Nozzle}
\end{figure}

We consider the acoustic response of a cylindrical 
nozzle flow. The flow is taken to be an inviscid, non-heat 
conducting perfect gas. The flow variables are assumed 
axisymmetric. The geometry of the nozzle,
showed in figure~\ref{fig:Mean flow Nozzle}, is characterised by a 
ratio between the cross-sectional areas of the upstream 
and downstream ducts of $\eta=25.$ The profile of the nozzle geometry 
is linear with a total length of $L/R_d=5.29$ (with $R_d$ the 
radius of the downstream duct). The upstream and downstream ducts 
are linked to the nozzle by two 
arcs of circumference of normalised radii $R/R_d=2.0.$  
The upstream and downstream ducts are $200 R_d$ long each. 
The geometry is discretised by a two-dimensional, fully non-structured 
mesh composed of $103,574$ triangles.

The mean flow is
computed using a steady-state solver for the 
non-linear compressible Euler equations in conservation 
variables. The equations are discretised in space using a 
continuous-Galerkin formulation stabilised 
using the least-squares method~\cite{donea2003finite}. 
The discretised non-linear problem is solved 
using a fully-implicit, pseudo-time-stepping
algorithm~\cite{crivellini2013spalart}. Second-order 
interpolation polynomials ($p=2$) were used for the computations.
The flow, which is fully defined by a Mach number 
at the inlet of $M_u = 0.0212,$ is accelerated at the nozzle to 
a Mach number of $M_d = 0.7029$ at the downstream duct, 
as observed in figure~\ref{fig:Mean flow Nozzle}.

\begin{table}
\begin{center}
\begin{tabular}{lcccc}
Forcing      &  $P_u^{-}$ (CAA) & $P_u^{-}$ (Model) & $P_d^{+}$ (CAA)  & $P_d^{+}$ (Model) \\
\hline
$P_u^+$      &  0.9764     &  0.9763  &  1.2752     &  1.2754 \\
$P_d^-$      &  0.0177     &  0.0177  & -0.1635     & -0.1632\\
$\sigma_u^+$ & -0.0073     & -0.0073  &  0.1401     &  0.1402\\
\end{tabular}
\end{center}
\caption{Validation of the inviscid part of the linearised 
Navier-Stokes equations (LNSE) solver. Reflection, $P_u^{-}$, and transmission, $P_d^{+},$ 
coefficients obtained numerically (CAA) and predictions of 
the compact Marble and Candel model~\cite{marble77}.}\label{table:Nozzle Predictions}
\end{table}

To compute the perturbations variables, both the molecular and turbulent viscosities
are set to zero in Eq.~\eqref{eq:Acoustic equations} and the coupled equations are solved
in cylindrical coordinates assuming axisymmetry of the flow variables. The equations 
are solved for a normalised frequency of $He=0.004$ and for three types of 
incoming excitations: (a) downstream-propagating acoustic waves at 
the inlet $P_u^+,$ (b) upstream-propagating acoustic waves at the 
outlet $P_d^-$ , and (c) entropy waves at the inlet $\sigma_u^+$. Again, 
second-order interpolation polynomials ($p=2$) were used for the computations.
The results are decomposed into upstream/downstream 
propagating components using the multi-microphone 
method in two post-processing zones spanning $-200 \leq x/R_d \leq -50$ 
and $50 \leq x/R_d \leq 200$ for the upstream and downstream ducts, respectively. The 
results obtained are summarised in table~\ref{table:Nozzle Predictions}, together with the 
predictions of the Marble and Candel model. The agreement is excellent for the three types of forcing, with the values being in agreement to, at least, the third decimal place. 

\section{Validation case: Acoustic response of the area expansion}\label{sec:Area expansion acoustic}
To further validate the code, we compare the numerical 
results obtained with the present implementation 
with the experimental dataset for an area 
expansion of Ronneberger~\cite{ronneberger1987}. This 
experiment has been used to benchmark several numerical 
approaches~\cite{kierkegaard2010frequency,gikadi2012linearized,foller2012identification,arina2016validation}.
Most of the parameters used in the experiment are the same as the ones used
throughout this paper: the expansion ratio, $\eta$, 
the working fluid and the Mach number at the inlet $M_u=0.2$. The 
Reynolds number of the experiment corresponds to $Re\approx 10^5$ which is 
lower than the value used in this publication $Re=10^6$.

 \begin{figure}
\centering
	\subfloat{\begin{tikzpicture}
				\node [inner sep=0pt,above right] 
                {\includegraphics[width=\textwidth]{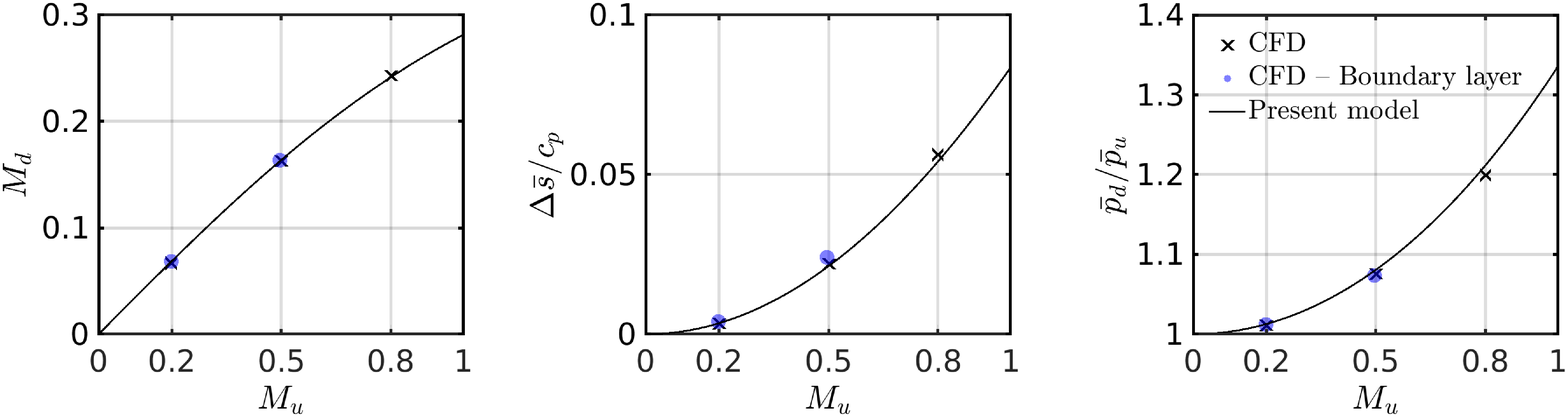}};
              \node at (0,3.7) {(a)};
			  \node at (4.2,3.7) {(b)};
              \node at (8.5,3.7) {(c)};
 			  \end{tikzpicture}} 
\caption{Computational results (CFD)
imposing (black crosses) a slip and (blue circles) 
a non-slip boundary condition at the upstream duct wall. (a) Downstream Mach 
number, $M_d,$ (b) entropy jump , $\Delta \bar{s}/c_p,$ 
and (c) mean pressure ratio, $\bar{p}_d/\bar{p}_u.$ }
\label{fig:Mean flow Model -- Boundary Layer}
\end{figure}

A fundamental difference of the results presented in this paper with the experimental
measurements is that the latter has a boundary layer at the wall of the upstream duct, 
while here we assume a uniform profile. In order to perform 
a fair comparison, an additional mean flow is computed 
prescribing a non-slip boundary 
condition at the wall of the upstream duct. At the inlet, a fully developed  
velocity profile is prescribed. Such profile was
obtained from an additional incompressible simulation of a uniform, straight duct.
The turbulent variables of the developed profile are also specified. Additionally, 
a zero-gradient is imposed to the pressure and a uniform profile to the temperature. 
The domain is the same described in Sec.~\ref{sec:mean flow numerical}. A 
fully structured mesh composed of $795,000$ cells is used here. Again, the viscous 
sublayer was resolved and, thus, the mesh was designed so that $y^+\sim 1$ 
is satisfied at the walls. Two Mach numbers were 
simulated: $M_u=0.2$ and $M_d=0.5.$
Figure~\ref{fig:Mean flow Model -- Boundary Layer} compares 
the results with the uniform case showing good agreement.

\begin{figure}
\centering
	\subfloat{\begin{tikzpicture}
				\node [inner sep=0pt,above right] 
                {\includegraphics[width=\textwidth]{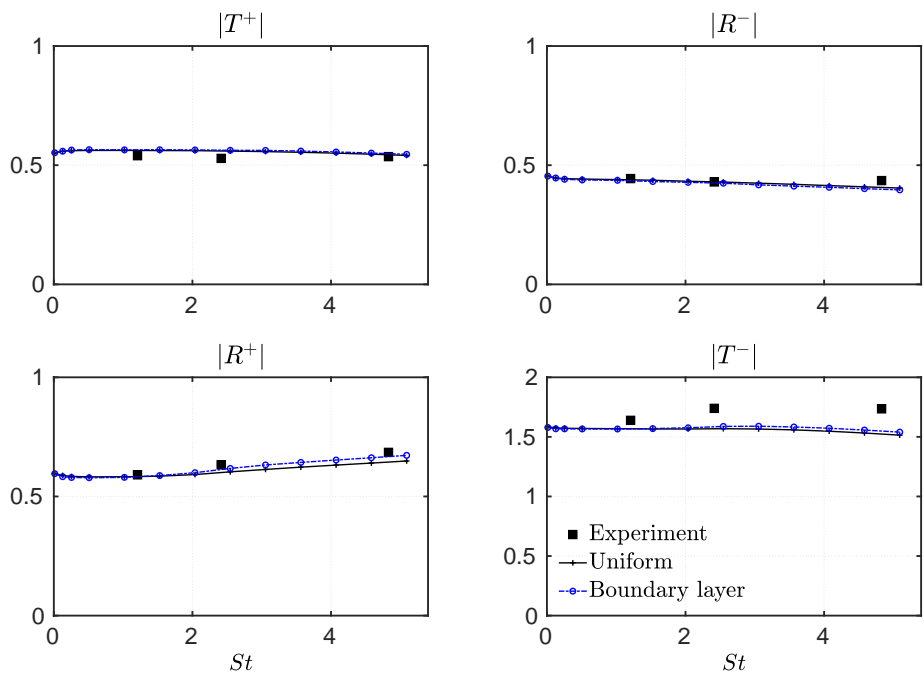}};
 			  \end{tikzpicture}} 
\caption{Magnitude of the scattering matrix at $M_u=0.2.$ Numerical results obtained 
with the coupled solver imposing (black solid line) a slip and
(blue dashed-dotted line) a non-slip boundary condition at the upstream 
duct for the mean flow. (Squares) experimental results.}
\label{fig:Acoustic Uniform Mu 0.2}
\end{figure}
The acoustic response is then computed using both mean flows (slip and 
non-slip upstream wall) for $M_u=0.2.$ The boundary conditions, meshes and numerical 
parameters are defined in Sec.~\ref{sec:entropy}. The simulations were performed using 
the coupled solver. The relations between acoustic 
waves are summarized in the scattering 
matrix~\citep{aabom1991measurement} which is defined as
\begin{equation}
\begin{pmatrix}
\hat{p}^+_d \\ \hat{p}^-_u 
\end{pmatrix}=
\begin{pmatrix}
T^+ & R^- \\
R^+ & T^-
\end{pmatrix}
\begin{pmatrix}
\hat{p}^+_u \\ \hat{p}^-_d
\end{pmatrix}.
\label{eq:scattering matrix}
\end{equation}
Figure~\ref{fig:Acoustic Uniform Mu 0.2} shows the results obtained for the scattering 
matrix. As can be seen, the simulated results agree well with experimental data using 
both mean flow fields. $T^{-}$ is underpredicted, but this trend is consistent
with previous numerical studies
\cite{kierkegaard2011simulations,foller2012identification,arina2016validation}. For higher
frequencies, the acoustic results obtained for separations with a uniform flow
and a flow with a boundary layer in the upstream duct diverge.   

\section{Numerical accuracy}\label{sec:numerical properties}
\begin{figure}
\centering
\includegraphics[width=0.6\textwidth]{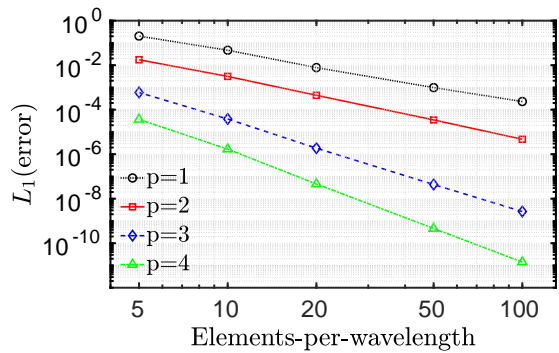}
\caption{Error between the analytical and numerical solutions obtained for an acoustic
 plane wave. $p$ is the degree of the interpolation polynomial.}
\label{fig:Numerical accuracy}
\end{figure}

In this appendix, we study the accuracy of the numerical 
implementation proposed in this paper for a given number 
of points per wavelength and order of the interpolation 
polynomials. To this end, we solve the perturbation
equations for the coupled formalism (Eq.~\eqref{eq:Acoustic equations}) in a 
fixed domain $\left[0,1\right]\times\left[0,1\right]$. 
The domain is discretised by five fully non-structured 
meshes composed of triangles. Each mesh has the same number of 
elements prescribed 
to each of its sides and this number is used 
to define the number of elements per wavelength
used in figure~\ref{fig:Numerical accuracy}. 

Non-reflecting boundary conditions for 
the acoustic waves are prescribed at 
left-hand and right-hand sides of the domain. Additionally, a 
left-to-right-propagating acoustic wave is introduced
at the left side. The variables 
are assumed axisymmetric, with the bottom side being 
the axis of revolution. At the top side, a slip boundary condition is imposed.
The mean flow is quiescent with constant and unitary speed of sound and 
mean density. The frequency is chosen to be $He=2\pi,$ selected so 
that one complete period is contained within the computational domain.

The L1-norm of the error for the acoustic pressure is evaluated as 
\begin{equation}
L_1(\text{error})= \frac{1}{N_{{\rm dof}}}\sum_{i=1}^{N_{{\rm dof}}}|\hat{p}_i - \hat{p}_{i, \text{exact}}|,
\label{eq:errr}
\end{equation} 
where $N_{{\rm dof}}$ is the number of degrees of freedom for 
each configuration. The exact solution to this problem is
$\hat{p}_{i,\text{exact}}=\exp\left(-\ii\, He\, x \right).$ 
Figure~\ref{fig:Numerical accuracy} shows that for the less resolved 
configuration presented in this paper (10 elements per wavelength
with second-order polynomials, $p=2$) the error per wavelength 
is $\mathcal{O}(10^{-3})$.  

\bibliography{sudden.bib}

\end{document}